\documentclass[fleqn,usenatbib]{mnras}

\usepackage{newtxtext,newtxmath}

\usepackage[T1]{fontenc}

\DeclareRobustCommand{\VAN}[3]{#2}
\let\VANthebibliography\thebibliography
\def\thebibliography{\DeclareRobustCommand{\VAN}[3]{##3}\VANthebibliography}

\usepackage{graphicx}	
\usepackage{amsmath}	

\title[New Method of Measuring Magnetic Field]{A New Method of Measuring Magnetic Field Strength in Highly Structured Protostellar Envelopes}

\author[Y. Tu et al.]{
Yisheng Tu,$^{1,2,3}$\thanks{E-mail: yitu@umich.edu}
Xiaoyuan Yang,$^{4}$
Zhi-Yun Li$^{2,3}$
\\
$^{1}$Department of Astronomy, University of Michigan, Ann Arbor, MI, 48109, USA\\
$^{2}$Astronomy Department, University of Virginia, Charlottesville, VA 22904, USA\\
$^{3}$Virginia Institute of Theoretical Astronomy, University of Virginia, Charlottesville, VA 22904, USA\\
$^{4}$Department of Physics, The Chinese University of Hong Kong, Shatin, N.T., Hong Kong, China
}

\date{Accepted XXX. Received YYY; in original form ZZZ}

\pubyear{\the\year{}}

\begin{document}
\label{firstpage}
\pagerange{\pageref{firstpage}--\pageref{lastpage}}
\maketitle

\begin{abstract}
Magnetic fields play a fundamental role in protostellar collapse and disk formation, yet direct measurements of magnetic field strength in deeply embedded protostellar envelopes remain difficult. We present a new method to estimate both the vertical and total magnetic field strength in collapsing, pseudodisk- or sheetlet-dominated protostellar envelopes, derived directly from the magnetohydrodynamic momentum equation. The method relates the magnetic field strength to two observationally accessible quantities: the projected gravitational acceleration ($g_R$) toward the center of collapse and the face-on column density ($\Sigma$) of the pseudodisk, and two dimensionless parameters, $\alpha_{b, R}$ and $\gamma_{zR}$, which characterize magnetic contribution to the force balance and the field geometry, respectively, through $|B_z|\approx(2\pi\alpha_{b,R}\gamma_{zR}g_R\Sigma)^{1/2}$. Using non-ideal magnetohydrodynamic simulations, we verify the assumptions underlying the method, justify the adopted approximations, and calibrate the two key dimensionless parameters. We provide canonical estimates of these two parameters, and show that they exhibit only weak spatial and temporal variations, allowing robust field strength estimates even when detailed gas kinematics or high-resolution polarization information is unavailable. We show that the method is applicable in both turbulent and non-turbulent envelopes and is insensitive to the ambipolar diffusion coefficient, making it robust against uncertainties in the local turbulence strength and ionization rate. We apply the method to the Class 0 source L1157, using column-density and gravitational-acceleration estimates from the literature to estimate the magnetic field strength for L1157. Our result is broadly consistent with previous estimates from independent methods, demonstrating the utility of this approach for constraining magnetic fields in embedded protostellar systems.
\end{abstract}




\section{Introduction}
Magnetic fields play a crucial role in star formation and have been detected across all evolutionary stages, from pre-stellar cores \citep[e.g.,][]{Torland2008, Kandori2018, Nakamura2019, Myers2021, Lin2024} to Class 0/I protostars \citep[][]{Lee2014b} and Class II disks \citep[][]{Ohashi2025}. At later stages, once a disk has formed, magnetic fields are thought to drive key processes in disk evolution: they launch outflows and jets that remove angular momentum \citep[e.g.,][]{Blandford1982, Shu1994, Machida2008, Zanni2013, Jacquemin2019, Jannaud2023, Zhu2025, Tu2025c, Tu2025b}, and they drive instabilities such as the magneto-rotational instability (MRI), which facilitates substructure formation within disks \citep[e.g.,][]{Machida2005, Bai2011, Suriano2017, Zhu2018, Delage2022, Hu2022, Zhu2024, Hsu2024}. The stage linking the early collapse of a cloud core to the later disk-dominated phase is a highly dynamic period during which the core is actively collapsing. Although this stage is relatively short compared to the long-lived Class II phase, the evolution of the magnetic field during this time is critical. It determines how much magnetic flux is retained in the forming disk, which in turn sets the course for the subsequent evolution of the disk and the eventual formation of planets.

It is well established that not all magnetic flux threading a cloud core can be transported into the disk \citep[e.g.,][]{Allen2003, Hennebelle2009}. If it were, the sharp concentration of magnetic fields would dramatically amplify their strength, and the resulting magnetic braking would be so efficient that disk formation would be suppressed entirely: a scenario known as the magnetic braking catastrophe \citep[][]{Mellon2008, Li2011}. This outcome, however, is in obvious contradiction with the widespread observations of disks around young stars \citep[e.g.,][]{ALMA2015}. To resolve this, magnetic fields must be redistributed during collapse, primarily through non-ideal MHD effects that decouple the gas from the field \citep[][]{Mellon2009, Li2011, Dapp2012, Wurster2018, Wurster2021, Xu2021, Tu2024a, Tu2024b}. This leads to a key question: how much magnetic flux is redistributed by non-ideal MHD processes within the protostellar envelope during the collapse stage?

In the absence of turbulence, the collapsing protostellar envelope forms the well-known ``pseudodisk'' \citep[][]{Li2014PPVI, Li2014b, Lee2014a, Lee2019}: a thin, flattened structure that lies in the same plane as the growing circumstellar disk. Because the pseudodisk is not rotationally supported against infall, it accretes rapidly toward the star, dragging magnetic fields inward and delivering both mass and magnetic flux to the disk. As a result, the field lines take on an ``hourglass'' morphology, as identified in dust polarization observations \citep[][]{Girart2006, Maury2018, Kwon2019, Choi2025} and appear as ``pinched'' field lines in numerical simulations \citep[][]{Lubow1995, Krasnopolsky2012, Machida2020, Wang2025}.

When the protostellar core is initially turbulent, as observations of pre-stellar cores suggest \citep[][]{Burkert2000, Chen2019, Ha2022}, the pseudodisk is instead distorted into a three-dimensional structure. \citet[][]{Tu2024a} identified these 3D counterparts of the pseudodisk and termed them ``gravo-magneto sheetlets''. Similar to pseudodisks, these sheetlets form through the combined action of gravity and magnetic fields and serve a similar role in transporting mass and magnetic flux into the disk. Although numerical modeling of these pseudodisk/sheetlets can, in principle, address the question of flux redistribution, large uncertainties in key parameters, particularly the local, non-equilibrium ionization state, make constructing a fully predictive model challenging. Observational constraints, whether direct or indirect measurements of magnetic fields on the pseudodisk/sheetlets, are therefore essential to guide and calibrate these theoretical efforts.

One key parameter that is notoriously difficult to measure observationally is the magnetic field strength. While the magnetic field topology--such as the characteristic ``hourglass'' morphology--can be inferred from polarization observations, determining the field strength itself generally requires much more subtle diagnostics. The most direct observational method for measuring magnetic field strength is the Zeeman effect, which provides a direct probe of the line-of-sight magnetic field component. But the Zeeman effect has yet to be definitively detected in the protostellar envelope of low-mass stars \citep[][]{Crutcher2019}. Similarly, the low ionization fraction in the protostellar envelope environment also makes measuring Faraday rotation challenging \citep[][]{Tahani2018}. A commonly used indirect way is by using the Davis-Chandrasekhar-Fermi (DCF) method \citep[][]{Davis1951, Chandrasekhar1953, Liu2021, Chen2022, Pattle2023}, where observable kinematics are modeled as perturbations of an ordered field. The DCF method needs to be used with great caution, since the conditions under which it was derived (such as small perturbations to an initially uniform magnetic field in a static background medium) are generally not satisfied in a protostellar envelope, where mass is moving rapidly towards a central gravitational source. Since then, other methods have been proposed, such as using dust polarization observations and inferred magnetic field vectors \citep[e.g.,][]{Koch2012, Shane2024}. A systematic way of estimating magnetic field strength, taking into account the structure of the protostellar envelope and validated with numerical simulation, is lacking. This will be the focus of this paper. 

The key conceptual advance of this work is that magnetic field strength can be inferred from force balance alone, without requiring explicit reconstruction of the magnetic field morphology. This makes the method applicable even in complex, asymmetric, or turbulent environments where traditional techniques become unreliable. Our work uses the numerical simulation presented in \citet[][]{Tu2024a} to justify and test our proposed method of estimating magnetic field strength. \citet{Tu2024a} performed a suite of numerical simulations examining the protostellar envelope structure with different turbulence strength and magnetic field diffusivity. Our paper is orgnaized as follows: We will derive and describe our method of estimating magnetic field strength in Section~\ref{sec:method_of_measure}, and explicitly list out the assumptions to be justified. We justify these assumptions and test the validity of our method using the models with initially laminar prestellar cores in section~\ref{sec:validation_noturb}. The method is then generalized to models with initially turbulent prestellar cores in Section~\ref{sec:validation_turb}. We discuss observational implications and connection to other methods in section~\ref{sec:discussion}. We summarize and conclude in section~\ref{sec:conclusion}.



\section{Method of measuring the magnetic field strength}
\label{sec:method_of_measure}
In this section, we describe the theoretical framework leading to our proposed method for measuring magnetic fields using kinetic information and polarization data. We will highlight our assumptions in each step and justify them one by one in the following subsections.

The forces in a magnetohydrodynamics system are described by the momentum equation
\begin{equation}
    \rho\frac{\partial \mathbfit{v}}{\partial t} + \rho(\mathbfit{v}\cdot\nabla)\mathbfit{v} - \frac{1}{4\pi}(\nabla\times\mathbfit{B})\times\mathbfit{B} + \nabla P - \rho\mathbfit{g} = 0,
    \label{equ:mhd momentum equation}
\end{equation}
where $\rho$ and $P$ are the density and pressure respectively; $\mathbfit{v}$, $\mathbfit{B}$, and $\mathbfit{g}$ are the velocity, magnetic field, and gravitational acceleration vectors respectively. The gravitational acceleration is defined as $\mathbfit{g}=-\nabla\Phi$, including both the gravity of the protostar and the gas self-gravity. Because the forces in the protostellar envelope are dominated by gravity, we can normalize equation~\ref{equ:mhd momentum equation} by gravitational force, and define
\begin{equation}
    \boldsymbol{\alpha}_t = \rho\frac{\partial \mathbfit{v}}{\partial t} / \rho\mathbfit{g};
    \label{equ:alpha_t}
\end{equation}
\begin{equation}
    \boldsymbol{\alpha}_v = \rho(\mathbfit{v}\cdot\nabla)\mathbfit{v} / \rho\mathbfit{g};
    \label{equ:alpha_v}
\end{equation}
\begin{equation}
    \boldsymbol{\alpha}_b = - \frac{1}{4\pi}(\nabla\times\mathbfit{B})\times\mathbfit{B} / \rho\mathbfit{g};
    \label{equ:alpha_b}
\end{equation}
\begin{equation}
    \boldsymbol{\alpha}_p = \nabla P / \rho\mathbfit{g};
    \label{equ:alpha_p}
\end{equation}
and by definition,
\begin{equation}
    \boldsymbol{\alpha}_t + \boldsymbol{\alpha}_v+\boldsymbol{\alpha}_b+\boldsymbol{\alpha}_p=\boldsymbol{1}
    \label{equ:mhd alpha equ}
\end{equation}
Physically, these $\alpha$ parameters quantify the fraction of the gravitational force that is counterbalanced by, or reacted to, by each force component. Since gravity is the primary driver of the dynamics during protostellar collapse, each force is naturally expected to scale as some fraction of the gravitational force.

We now describe and derive our method for estimating the magnetic field strength, explicitly stating each assumption involved. These assumptions are subsequently justified one by one in Section~\ref{sec:validation_noturb}.

\textbf{Assumption 1}: The time dependence term contributes minimally to the total force balance. 

\textbf{Assumption 2}: The pressure gradient term also contributes minimally to the total force balance.

These two assumptions allow us to rearrange equation~\ref{equ:mhd alpha equ} to
\begin{equation}
    \boldsymbol{\alpha}_v + \boldsymbol{\alpha}_b = (1 - \boldsymbol{\alpha}_t -\boldsymbol{\alpha}_p)
    \label{equ:mom_equ_simp_01}
\end{equation}

\begin{figure*}
    \centering
    \includegraphics[width=\linewidth]{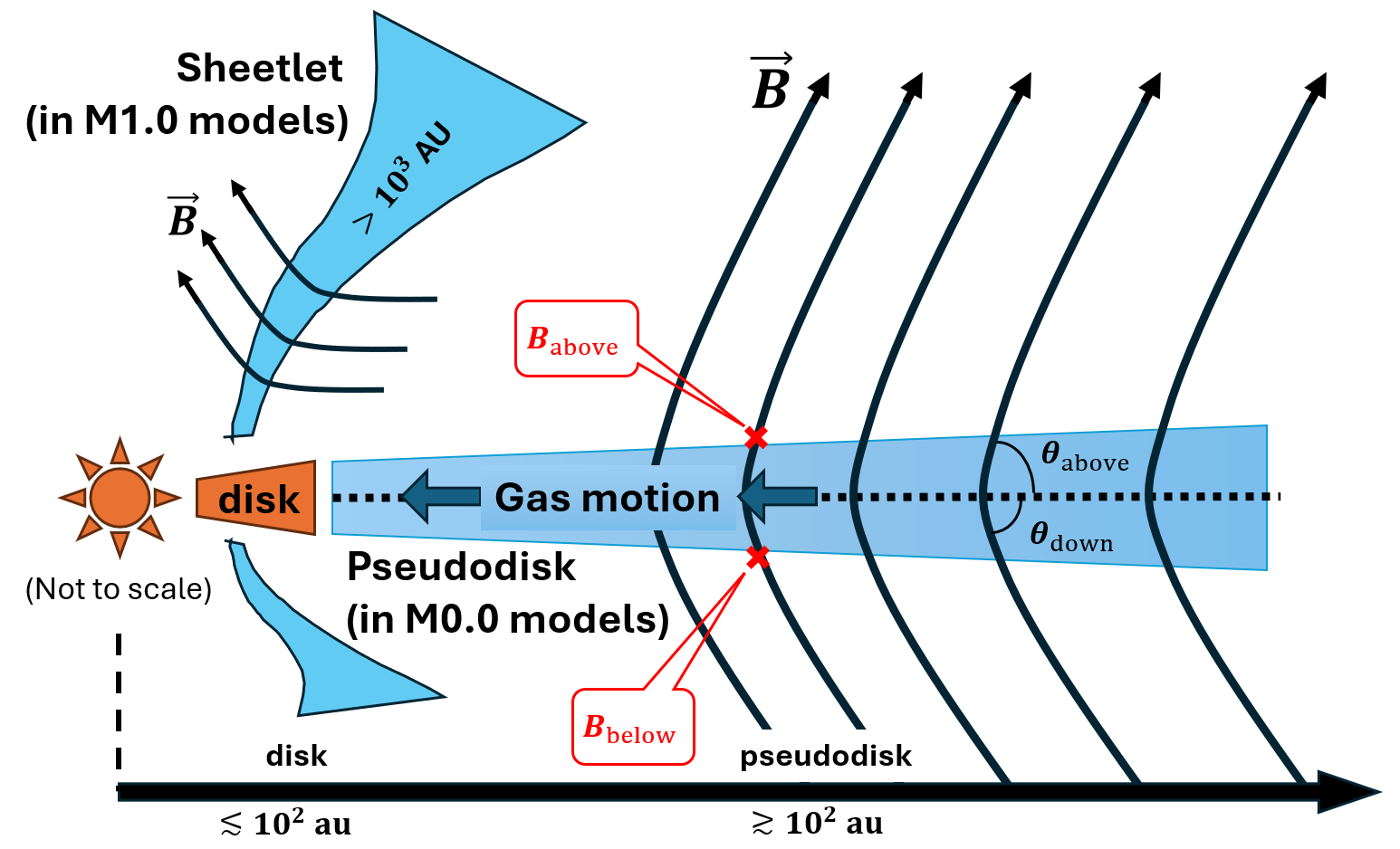}
    \caption{Cartoon of the envelope structure in a collapsing magnetized protostellar system. The center of the system is the star and its circumstellar disk; In the models where the core is initially laminar (i.e., the M0.0 models), the protostellar envelope is dominated by the ``pseudodisk''; in the initially turbulent core models (i.e., the M1.0 models), the protostellar envelope is dominated by the ``gravo-magneto-sheetlets'' \citep[see][for a more detailed description]{Tu2024a}. Both the pseudodisk and sheetlets dominate mass and magnetic field transport within the protostellar envelope, and our goal here is to estimate the magnetic field strength on these pseudodisk/sheetlets.}
    \label{fig:sketch}
\end{figure*}

\textbf{Assumption 3}: The magnetic force term is dominated by the magnetic tension force along a pseudodisk \citep[or sheetlet as their turbulent counterpart, see figure~3 in][hereafter, we will define pseudodisk as a thin structure on the equatorial plane, whereas sheetlets exhibit a 3D geometry]{Tu2024a}:
\begin{equation}
    \frac{1}{4\pi}(\nabla\times\mathbfit{B})\times\mathbfit{B} \approx \frac{1}{4\pi}(\mathbfit{B}\cdot\nabla)\mathbfit{B}.
    \label{equ:mom_equ_simp_tens}
\end{equation}
This assumption allows us to approximate the magnetic force using magnetic field geometry. Fig.~\ref{fig:sketch} shows a sketch of the approximate magnetic field geometry along a meridian slice through a pseudodisk in the non-turbulent models (M0.0 models), and the sheetlet in the turbulent models (M1.0 models, \citealt[][]{Tu2024a}). 

The sheetlets are relatively thin, so the gravitational force, which is mainly due to the central protostar in the inner envelope under consideration, does not vary much within them. We can use this thin geometry by integrating both the magnetic force and gravitational force vertically along the sheetlets, using a constant $\alpha_b$ (equation~\ref{equ:alpha_b}), i.e.,
\begin{equation}
    \int_\mathrm{sh}\frac{1}{4\pi}(\mathbfit{B}\cdot\nabla)\mathbfit{B}~dl\approx \int_\mathrm{sh} \alpha_b \rho \mathbfit{g} dl=\alpha_b\mathbfit{g}\int_\mathrm{sh}\rho dl=\alpha_b\mathbfit{g}\Sigma,
    \label{equ:alpha_b_integral}
\end{equation}
where the integration bounds (denoted by $\mathrm{sh}$) is over the height of the sheetlets; $\Sigma$ is the column density of the sheetlets.

The magnetic field on either side of the sheetlets is dominated by two components: One component perpendicular to the sheetlet (denoted as $B_\perp$), and the other component parallel to the direction of motion of the sheetlet ($B_\parallel$). In the case where the sheetlets lie on the equatorial plane (as in the case of the pseudodisk), the magnetic field can be written as: $\mathbfit{B}=\langle B_R, B_\phi, B_z\rangle$ in cylindrical coordinates, where $B_\perp=B_z$ and $B_\parallel=B_R$. We neglect $B_\phi$ in subsequent discussion as it contributes minimally to the force balance in the cylindrical radial direction, and its magnitude on the pseudodisk is small compared to $B_z$ and $B_R$ (see appendix~\ref{app:gamma_zphi}).

Given the magnetic field geometry is near symmetric about the sheetlets (i.e., $B_{\parallel, \mathrm{above}}\approx B_{\parallel, \mathrm{below}}\equiv B_\parallel$, see Figure~\ref{fig:sketch}), we can write the integrated magnetic force as the magnetic stress across the sheetlet
\begin{equation}
    \int_\mathrm{sh}\frac{1}{4\pi}[(\mathbfit{B}\cdot\nabla)\mathbfit{B}]\cdot\hat{s}~dl\approx \frac{B_\perp B_\parallel}{2\pi},
    \label{equ:mom_equ_simp_mag_final}
\end{equation}
where $\hat{s}$ is taken to be along the primary direction of motion of the sheetlets, and $\hat{s}=\hat{R}$, the cylindrical-$\hat{R}$ in the case of a pseudodisk; $dl$ is perpendicular to $\hat{s}$.
Note that $B_\perp^s$ and $B_\parallel^s$ are related through geometry: $B_\perp^s = B_\parallel^s\tan(\theta)$, where $\theta$ is the pinching angle of the magnetic field (Figure~\ref{fig:sketch}).

Combining the terms in $\hat{s}$ direction of equation~\ref{equ:alpha_b}, \ref{equ:mom_equ_simp_tens}, \ref{equ:alpha_b_integral} and \ref{equ:mom_equ_simp_mag_final}, we can estimate the magnetic field strength using

\begin{equation}
    B_\perp B_\parallel = \alpha_{b, s} 2\pi g_s \Sigma,
    \label{equ:est_B_final}
\end{equation}
where $\alpha_{b, s}$ is defined by equation~\ref{equ:alpha_b}, and $g_s$ is gravitational acceleration in the $\hat{s}$ direction respectively. 

Finally, if the magnetic field geometry can be measured and the pinching angle $\theta$ can be obtained, then we can obtain the magnetic field strength on the sheetlets. Alternatively, we can parameterize the geometry term as $\gamma_{\perp\parallel}\equiv\tan(\theta)$\footnote{In more laminar systems, $\gamma_{zR}$ reflects the geometric ratio of magnetic field components on the equatorial pseudodisk. However, in highly turbulent systems, this parameter may instead be interpreted as an effective or empirical correction factor that encapsulates complex field structure and fluctuations. See Section~\ref{sec:validation_turb} for a detailed discussion.}, so the perpendicular component of the magnetic field can be measured as
\begin{equation}
    B_\mathrm{\perp}^2=\gamma_{\perp\parallel}~\alpha_{b, s}~2\pi g_s\Sigma
    \label{equ:est_Bperp}
\end{equation}
or, equivalently
\begin{equation}
    |B_\mathrm{\perp}|=(\gamma_{\perp\parallel}~\alpha_{b, s}~2\pi g_s\Sigma)^{1/2}
    \label{equ:est_Bperp_nosq}
\end{equation}
In the following subsection, we justify the assumptions underlying this approach and demonstrate that equation~\ref{equ:est_Bperp} and \ref{equ:est_Bperp_nosq} can be used to estimate the magnetic field strength with reasonable accuracy using the simulation data. We will further provide canonical values for $\gamma_{\perp\parallel}$ and $\alpha_{b, s}$, and discuss their validity and limitations.

\section{Validation with numerical simulation: non-turbulent models}
\label{sec:validation_noturb}
In this section, we validate our method using numerical simulations. The simulations are taken from \citet[][]{Tu2024a}, who modeled the collapse of a molecular cloud core leading to the formation of a star and its circumstellar disk. We refer the reader to \citet[][]{Tu2024a} for a detailed description of the numerical setup, and summarize here the essential aspects of the physical setup and main results.

The suite of non-ideal magnetohydrodynamic (MHD) simulations was performed with the \texttt{ATHENA++} framework \citep[][]{Stone2020, Tomida2023}. The models include gas self-gravity, magnetic fields, and ambipolar diffusion—the dominant non-ideal MHD effect in protostellar envelopes. Two parameters distinguish the different models, and the model names reflect these choices. The first is the presence or absence of an initial turbulent velocity field, representing turbulence inherited from larger-scale molecular clouds. Its strength is quantified by the root-mean-square Mach number ($\mathcal{M}$). The second is the adopted non-ideal MHD coefficient ($\eta$), which sets the coupling strength between gas and magnetic fields. The reference value for the ambipolar diffusion coefficient ($\eta_0$) corresponds to a cosmic ray ionization rate of $10^{-17}~\mathrm{s}^{-1}$ \citep[][]{Shu1991}. A model name such as M1.0AD1.0 indicates an initial Mach number of 1.0 and an ambipolar diffusion coefficient equal to the reference value.

The models focus on low-mass star-forming dense cores with a total mass of $0.5,M_\odot$. The initial condition is a supercritical pseudo–Bonnor-Ebert sphere with a radius of 2000 au, threaded by vertical magnetic fields. Core collapse leads to the formation of a central star and its circumstellar disk. \citet[][]{Tu2024a} found that the envelope surrounding the forming star and disk is highly structured, dominated by the pseudodisk if the core is initially laminar \citep[see also,][]{Galli1993, Allen2003, Machida2019}, and by the ``gravo-magneto-sheetlets'' (or simply ``sheetlets'') if the core is initially turbulent \citep[see Figure 3 of][for a 3D visualization]{Tu2024a}. Both the pseudodisk and the sheetlets share the same physical origin, as both arise from the interplay between gravity and magnetic fields. Magnetic field lines are sharply pinched on these pseudodisks/sheetlets, slowing infall and extracting angular momentum from the accreting material. To identify the pseudodisks/sheetlets in our simulation, we follow the definition in \citet[][]{Tu2024a}, and define the pseudodisks/sheetlets that satisfy the following three criteria: 1) $\rho<5\times 10^{-15}~\mathrm{g~cm^{-3}}$; 2) above the density threshold $\rho_c$ that contains 70\% of the mass; and 3) plasma-$\beta>1$.

To validate our method for measuring magnetic field strength in simulations, we begin with the simplest case: a model without initial turbulence. In this setup, the sheetlets reduce to the well-known pseudodisk that lies on the equatorial plane. The flat geometry of the pseudodisk allows us to define $B_\perp$, $B_\parallel$, $\Sigma$, $g_s$ and $\alpha_{b, s}$ in equation~\ref{equ:est_B_final} using cylindrical coordinates, where the $\hat{z}$-axis is perpendicular to the disk plane: $B_\perp = B_z$, $B_\parallel = B_R$, $\Sigma$ is the pseudodisk column density, $g_s = g_R$ is the gravitational acceleration in the cylindrical $\hat{R}$-direction; and $\alpha_{b, s}=\alpha_{b, R}$. In this case, the magnetic field strength can be estimated as
\begin{equation}
    B_z B_R = \alpha_{b, R} 2\pi g_R \Sigma.
    \label{equ:BzBR}
\end{equation}
And the vertical magnetic field strength, which determines the amount of magnetic flux carried to the disk, can thus be estimated as
\begin{equation}
    B_z^2 = \gamma_{zR}\alpha_{b, R} 2\pi g_R \Sigma; \quad\mathrm{or}\quad |B_z| = (\gamma_{zR}\alpha_{b, R} 2\pi g_R \Sigma)^{1/2},
    \label{equ:estBz_M0.0}
\end{equation}
where the geometric parameter $\gamma_{\perp\parallel}=\gamma_{zR}$.

To validate our method under different ionization conditions, we present two models, M0.0AD1.0 \citep[][]{Tu2024a} and M0.0AD2.0 (appendix~\ref{app:M0.0AD2.0}), corresponding to a factor of 4 difference in the ionization rate. In the following subsections, we will validate the assumptions and estimate $\alpha_{b, R}$ and $\gamma_{zR}$ in both models.

\subsection{Assumption 1 \& 2: time dependence and pressure gradient term}
\label{sec:assumption12}
We start by showing that the time dependence and pressure gradient contribute minimally to the total force balance in pseudodisks in protostellar envelopes in the non-turbulent models. Since the time dependence term (equation~\ref{equ:alpha_t}) is hard to directly estimate with limited output frequency, we estimate its value by inverting the momentum equation (equation~\ref{equ:mhd alpha equ}) in the cylindrical-$\hat{R}$ direction: $[\alpha_t]_R=1-[\alpha_v+\alpha_p+\alpha_b]_R$. Since the pseudodisk lies on the midplane in these non-turbulent models, the cylindrical-$\hat{R}$ direction coincides with the primary flow direction and gravitational force direction on the pseudodisks. 

\begin{figure*}
    \centering  \includegraphics[width=\linewidth]{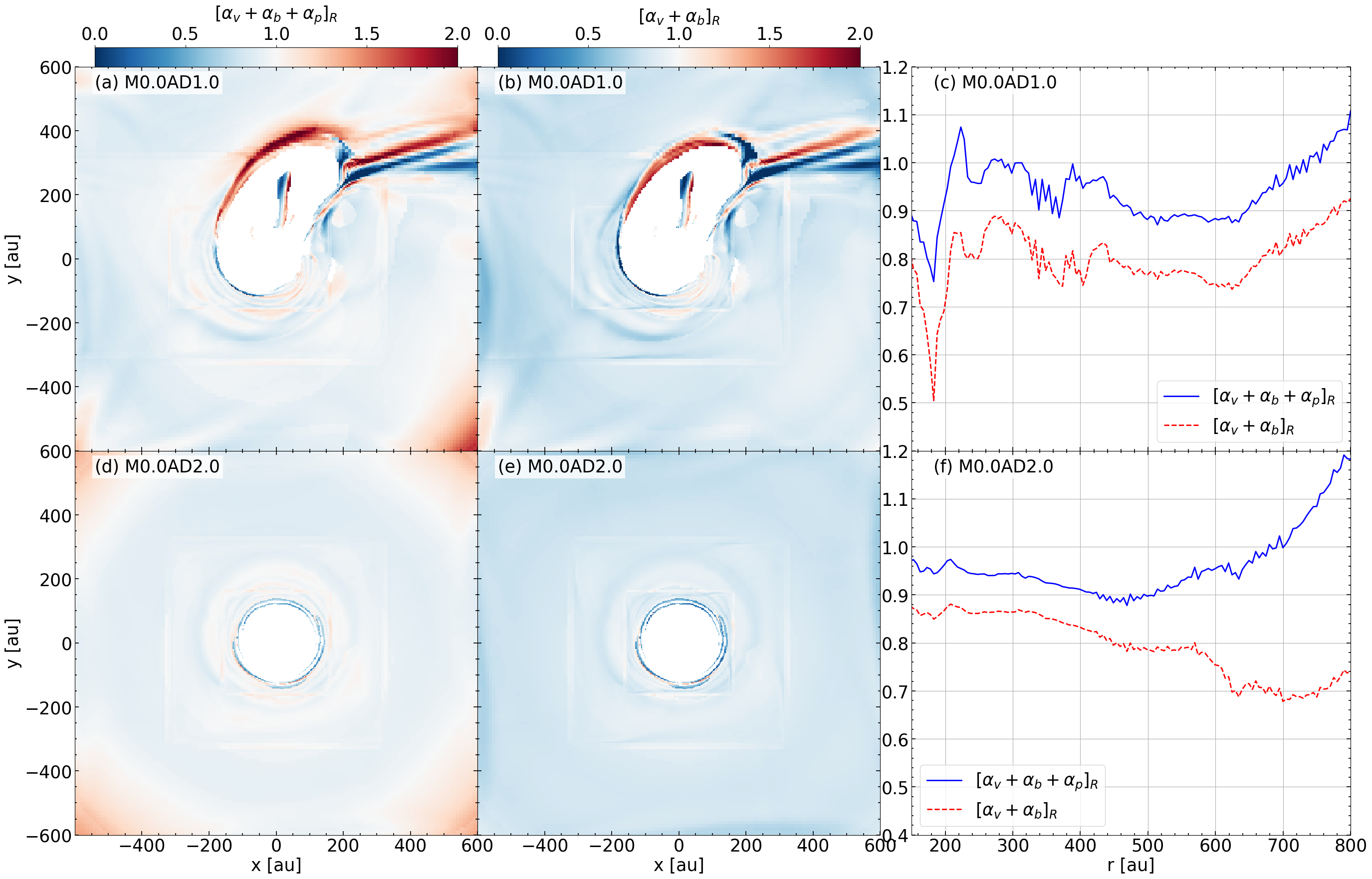}
    \caption{Force balance in the M0.0AD1.0 model when $M_\star=0.2M_\odot$ (40\% of total core mass). \textbf{Panel (a)}: Force balance on the pseudodisk excluding the time-dependence term (see equation~\ref{equ:alpha_t}- \ref{equ:mhd alpha equ}). The notations here only show the physical meaning of this figure, and the averaging is defined by equation~\ref{equ:weight_definition}. \textbf{Panel (b)}: Force balance on the pseudodisk excluding both the time-dependence and the pressure gradient terms. \textbf{Panel (c)}: azimuthally-averaged force balances in each annulus, excluding the time-dependence term only (blue line), and excluding both the time-dependence and pressure gradient terms (red line).}
    \label{fig:tot_force_balance_M0.0AD1.0}
\end{figure*}

Figure \ref{fig:tot_force_balance_M0.0AD1.0} presents the force-balance ratios at a representative time when the stellar mass is about 0.2$M_\odot$ (40\% of total core mass). To estimate the averaged force balance across the pseudodisk (see section~\ref{sec:validation_noturb}), we first integrate the relevant forces over the region of interest and then compute their ratio. Specifically, the averaged force-balance ratio is defined as (for example)
\begin{equation}
    [\alpha_v+\alpha_p+\alpha_b]_R \equiv \frac{\int_\mathrm{cells}[\rho \mathbfit{v}\cdot\nabla \mathbfit{v}+\nabla P-\frac{1}{4\pi}(\nabla\times \mathbfit{B})\times \mathbfit{B}]_R dV}{\int_\mathrm{cells}[\rho \mathbfit{g}]_RdV}
    \label{equ:weight_definition}
\end{equation}
where the integration is over a column of cells in figure~\ref{fig:tot_force_balance_M0.0AD1.0}(a) and (b), and over an annulus in figure~\ref{fig:tot_force_balance_M0.0AD1.0}(c). Hereafter, the force ratios are calculated using similar definitions.

Figure \ref{fig:tot_force_balance_M0.0AD1.0}(a) shows the column-weighted sum of $[\alpha_v + \alpha_p + \alpha_b]_R$ across the pseudodisk in the M0.0AD1.0 model. The central white region contains the disk and the magnetic bubble created by interchange instability \citep[similarly identified in e.g.,][]{Krasnopolsky2012, Machida2025}, neither of which is part of the pseudodisk. Across most of the pseudodisk, the sum remains slightly below but close to unity, indicating that the radial force balance is largely satisfied without including the time-dependence term.

Figure \ref{fig:tot_force_balance_M0.0AD1.0}(c) quantifies this by showing the azimuthally-averaged $[\alpha_v + \alpha_p + \alpha_b]_R$ within the pseudodisk as a function of spherical radius from the star (located at $r=0$). The average is computed using equation \ref{equ:weight_definition}, integrating over all pseudodisk cells within each radial annulus. This ratio stays near $\sim 1$ inside $\sim 400$ au, decreases to $\sim 0.9$ beyond 400~au.

The region inside 400 au contains a sharp structural boundary, the edge of the magnetic bubble, which introduces abrupt variations in the gas properties and elevates the apparent force-balance ratio (see figure~\ref{fig:tot_force_balance_M0.0AD1.0}a). Focusing on the bulk region of the pseudodisk, which lies primarily beyond 400 au, we therefore adopt a representative force-balance ratio (excluding the time-dependence term) of approximately 0.9; that is, $[\alpha_v + \alpha_p + \alpha_b]_R \approx 0.9$.

Similarly, figure~\ref{fig:tot_force_balance_M0.0AD1.0}(b) and (c) show the spatial and averaged force balance of (inertia + magnetic) force with gravity (i.e., $[\alpha_v + \alpha_b]_R$). The central $\sim150$ au is dominated by the disk and is therefore excluded from this analysis. Since pressure force is not included, the values are slightly less than those in figure~\ref{fig:tot_force_balance_M0.0AD1.0}(a) as expected, yet staying $\approx0.8$ at almost all radii. This well-balanced force, by including only the gas inertia and magnetic terms, shows that the total contribution of the time-dependence and pressure gradient terms is about 20\% (i.e., $[\alpha_t + \alpha_p]_R\approx 0.2$). The time-dependence and pressure gradient terms contribute similarly to the total force balance at other times in the simulation.

The ratios exhibit similar behavior in the more magnetically diffusive M0.0AD2.0 model (with a factor of 2 larger diffusivity, corresponding to a factor of 4 lower ionization level, compared to the reference model M0.0AD1.0), as shown in Figure~\ref{fig:tot_force_balance_M0.0AD1.0}(d-f), evaluated at a representative time when the stellar mass reaches $\sim 0.2~M_\odot$. As in the M0.0AD1.0 model, the total force-balance ratio remains close to unity, with $[\alpha_v+\alpha_p+\alpha_b]_R \approx 0.9$ over most radii within the pseudodisk. The ratio rises at large radii owing to the sharp gradients at the pseudodisk boundary. This again implies that $[\alpha_t]_R \approx 0.1$.

Similarly, the sum of the velocity and magnetic contributions satisfies $[\alpha_v+\alpha_b]_R \approx 0.8$, indicating that the combined time-dependence and pressure terms contribute only modestly, with $[\alpha_t+\alpha_p]_R \approx 0.2$ across the pseudodisk. The agreement between the M0.0AD1.0 and M0.0AD2.0 models demonstrates that these conclusions are robust against changes in ambipolar diffusivity: in all cases, both the turbulent and thermal pressure terms are subdominant, with typical values $[\alpha_t]_R \approx [\alpha_p]_R \approx 0.1$.

\subsection{Assumption 3: magnetic tension}
\label{sec:tension}
The next assumption we will justify is that the magnetic force is dominated by the magnetic tension force. To show this, we rearrange equation~\ref{equ:mom_equ_simp_tens}, and define
\begin{equation}
    \mathcal{F}_\mathrm{tens, R} = \Big[\frac{(\mathbfit{B}\cdot\nabla)\mathbfit{B}}{(\nabla\times\mathbfit{B})\times\mathbfit{B}}\Big]_R.
\end{equation}
where $\mathcal{F}_\mathrm{tens, R}$ is the contribution of the tension force in the cylindrical-$\hat{R}$ direction. 

\begin{figure*}
    \centering
    \includegraphics[width=\linewidth]{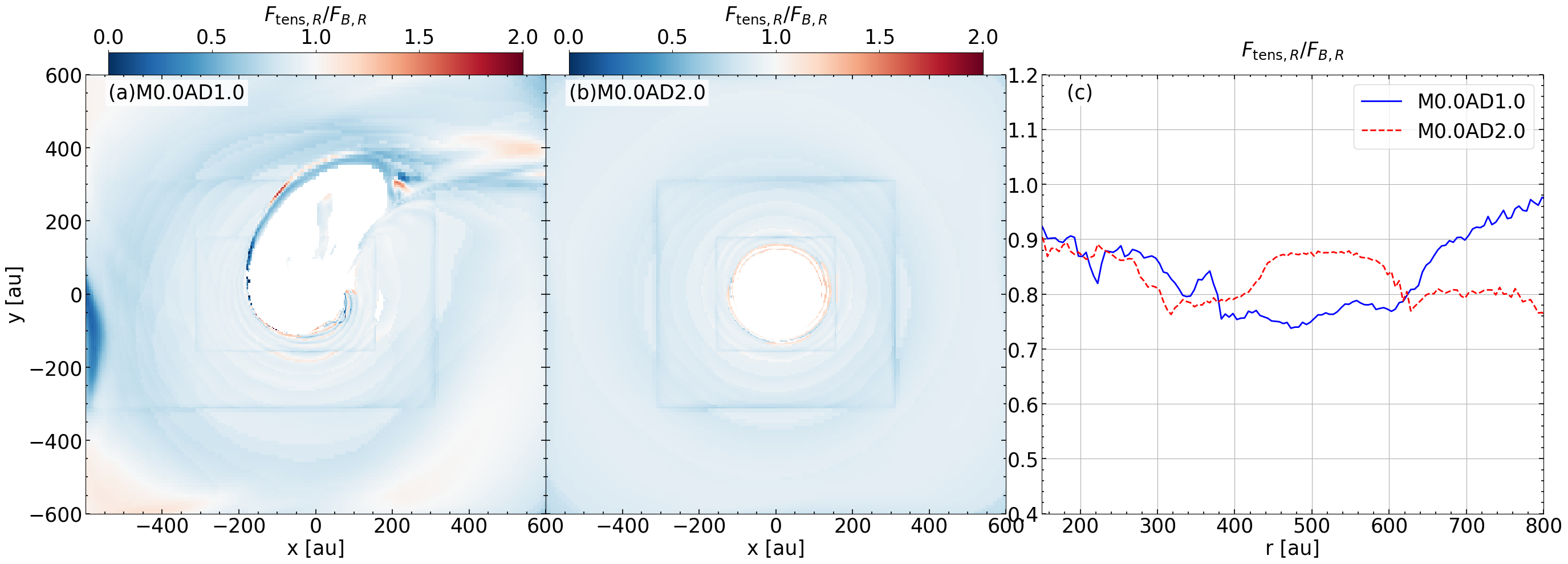}
    \caption{Ratio between the magnetic tension force and the total magnetic force in the cylindrical-$\hat{R}$ direction. {\bf Panel (a)} shows the ratio on the pseudodisk in the M0.0AD1.0 model; {\bf Panel (b)} shows the ratio in the M0.0AD2.0 model, and {\bf panel (c)} shows the azimuthal-averaged ratio in both models, indicating the total magnetic force can be approximated well with the magnetic tension force alone.}
    \label{fig:tension}
\end{figure*}

Figure~\ref{fig:tension} shows ratio between the cylindrical-$\hat{R}$ direction magnetic tension force ($F_\mathrm{tens, R}$) and the total magnetic force ($F_{B, R}$). The column-weighted ratios in the M0.0AD1.0 and M0.0AD2.0 models are shown in panels (a) and (b), respectively; the annuli-weighted ratios for both models are shown in panel (c). Both column- and annuli- averaging are done similarly as in equation~\ref{equ:weight_definition}, where the numerator and denominator are integrated separately. The values in both models are largely similar, holding $\mathcal{F}_\mathrm{tens, R}\gtrsim0.8$ at all radii and azimuthal angles.

\subsection{Measuring $\alpha_{b, R}$}
\label{sec:est_alpha_b_direct}
With assumption 3 justified, we are now ready to measure $\alpha_{b, R}$, the ratio between the magnetic force and the gravitational force in the cylindrical-$\hat{R}$ direction. In this subsection, we measure $\alpha_{b, R}$ through its definition, and in the following subsection (sec.~\ref{sec:est_alpha_b_alpha_v}), we estimate $\alpha_{b, R}$ through the kinematic information ($\alpha_{v, R}$).

There are two methods to measure $\alpha_b$: method 1 is by measuring equation~\ref {equ:alpha_b}, directly through definition; method 2 is through equation~\ref {equ:BzBR}, where we use the assumption that the magnetic force is dominated by the tension force (assumption 3). To fully take into account the field geometry on the pseudodisk, equation~\ref {equ:BzBR} can be rearranged as
\begin{equation}
    \alpha_{b, R}\approx\frac{(B_zB_R)_\mathrm{above} - (B_zB_R)_\mathrm{below}}{4\pi g_R \Sigma},
    \label{equ:alpha_b_uplow}
\end{equation}
where $(B_zB_R)_\mathrm{above}$ and $(B_zB_R)_\mathrm{below}$ are the values immediately above and below the pseudodisk and have opposite signs (see Figure~\ref{fig:sketch}). 

\begin{figure*}
    \centering
    \includegraphics[width=\linewidth]{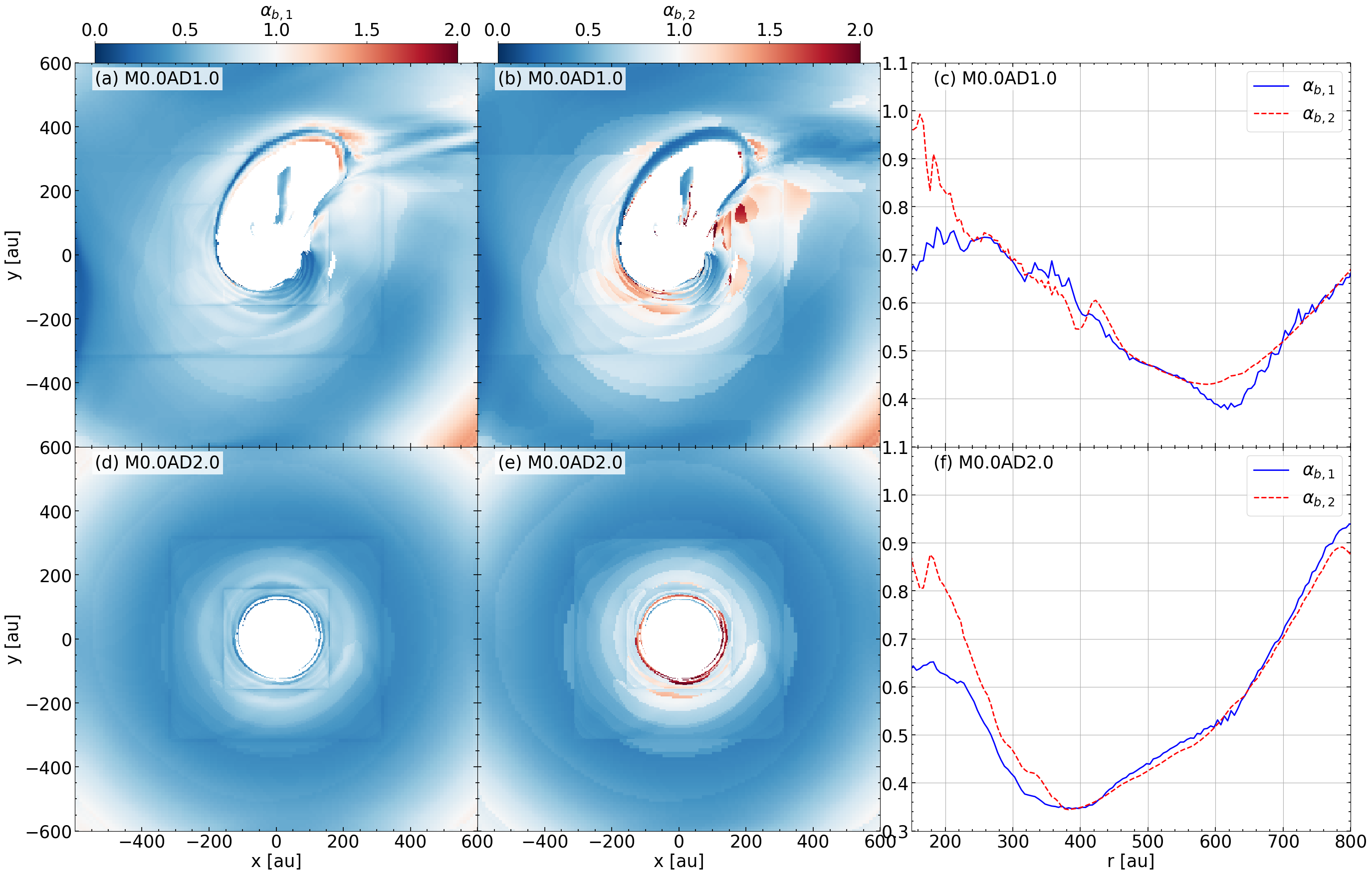}
    \caption{Measurement of $\alpha_{b, R}$ in the M0.0AD1.0 model in the upper panels and the M0.0AD2.0 model in the lower panels. Panel (a) and (d) show $\alpha_b$ through its definition (equation~\ref{equ:alpha_b}, method 1 in section~\ref{sec:est_alpha_b_direct}); Panel (b) and (e) show $\alpha_b$ through an approximation using the tension assumption and column density (equation~\ref{equ:alpha_b_uplow}, method 2 in section~\ref{sec:est_alpha_b_direct}). Panels (c) and (f) show their azimuthal average and a comparison between these two methods, revealing remarkable agreements. There is a characteristic ``V'' shape in Panels (c) and (f) due to a change in the magnetic field geometry responding to gas dynamics, as discussed in section~\ref{sec:est_alpha_b_direct}. An animated version of this figure is available at \url{https://doi.org/10.6084/m9.figshare.31445326}.}
    \label{fig:alpha_b_measure_direct}
\end{figure*}

Figure~\ref{fig:alpha_b_measure_direct}(a-c) show $\alpha_{b, R}$ measured by method 1 ($\alpha_{b, 1}$, panel a), and method 2 ($\alpha_{b, 2}$, panel b) in the M0.0AD1.0 model. These two methods show remarkable agreement with each other, indicating equation~\ref{equ:alpha_b_uplow} provides a very good estimation of the actual value (equation~\ref{equ:alpha_b}). Crucially, the magnetic force remains dynamically important across all radii: $\alpha_{b, R}$ is about 0.7 around 200 au, indicating magnetic force balances $70\%$ of the gravitational force; this ratio drops to about $40\%$ around 600 au, and rises back to $\sim 60\%$ at about 800 au. This shows the significance of the magnetic force in retarding the pseudodisk against infall in the protostellar envelope.

The ratio in the M0.0AD2.0 model, as shown in Figure~\ref{fig:alpha_b_measure_direct}(d), (e), and (f), shows a similar agreement and result. The range of $\alpha_{b, R}$ stays between 0.7 and 0.3, indicating a significant and comparable magnetic contribution in the model with higher magnetic diffusivity. Interestingly, this ratio shows a ``V-shaped'' trend as a function of radius: the value reaches a maximum around 0.7 around 150 au, then decreases with radius until $\sim 350$ au, where the value reaches a minimum of 0.3, and then gradually goes back up to $\approx0.8$ at about $800$ au. 

This ``V-shaped'' trend is also seen in the M0.0AD1.0 model at earlier times (see the animated version of Figure~\ref{fig:alpha_b_measure_direct}). The main driver of the outer part of this trend, where $\alpha_{b, R}$ increases towards larger radius (the outer part of the ``V''), is a delayed magnetic response to infall. At large radii, the collapse speed is low (small $v_r$), giving the magnetic field sufficient time to self-adjust and remain relatively straight. In this regime, magnetic tension, the dominant magnetic force in the pseudodisk (Section~\ref{sec:tension}), is weak. As infall accelerates, gravity grows rapidly, whereas the magnetic (tension) force does not, leading to a reduced magnetic contribution to the radial force balance (the outer part of the ``V'' shape, where $\alpha_{b, R}$ decreases with cylindrical radius).

\begin{figure}
    \centering
    \includegraphics[width=\linewidth]{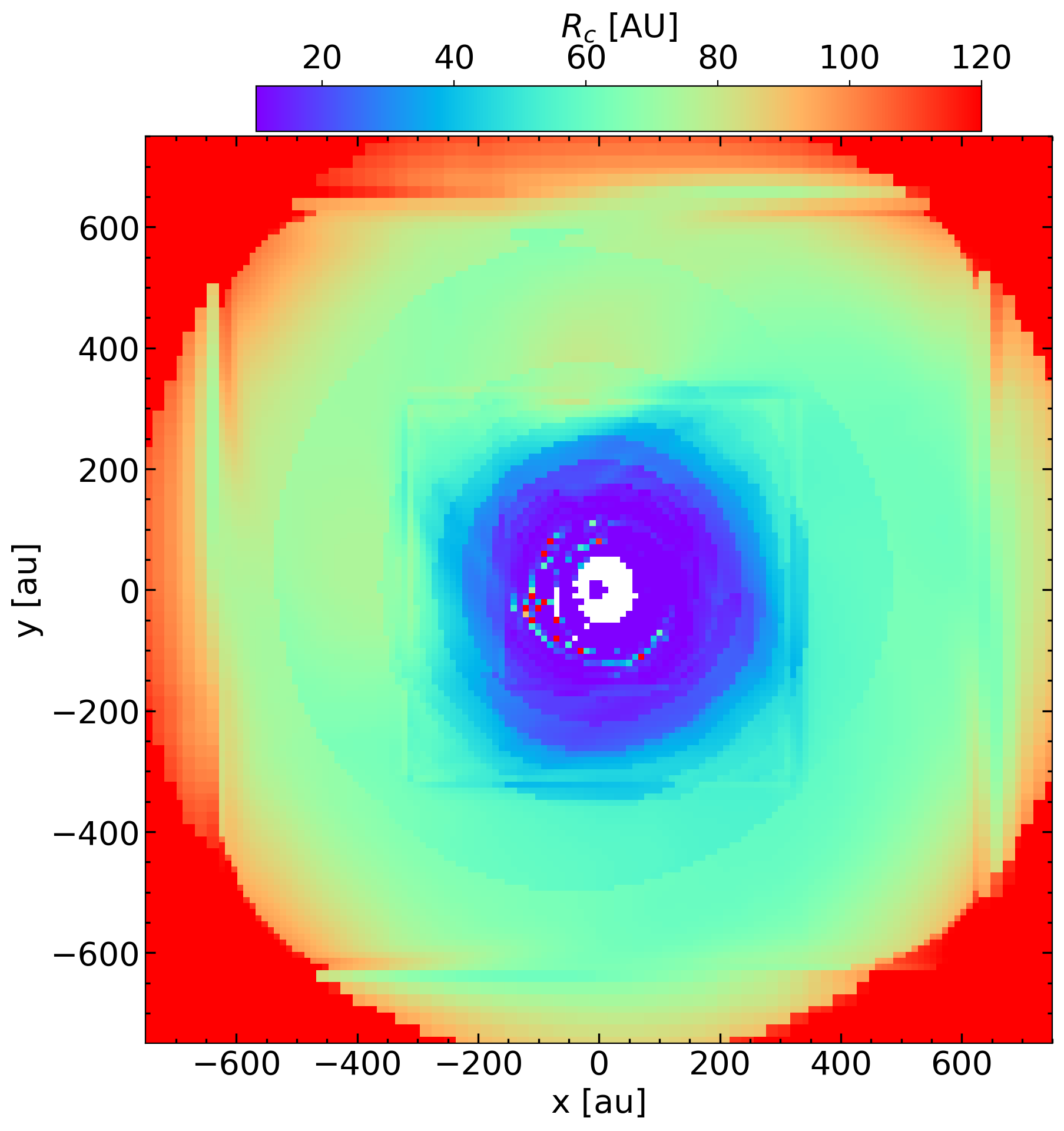}
    \caption{Illustration of magnetic field geometry using the estimated curvature radius of magnetic field ($R_c$, equation~\ref{equ:Rc}) in the M0.0AD2.0 model. A large $R_c$ indicates the field is relatively straight, whereas a small $R_c$ indicates the field is highly pinched. At large radii, outside the pseudodisk ($\gtrsim 600~\mathrm{au}$), $R_c$ is very large as the field is straight. In the intermediate radii on the pseudodisk, $R_c$ stays around $70~\mathrm{au}$. At small radii $(\approx200~\mathrm{au}$), $R_c$ decreases sharply to $\lesssim15~\mathrm{au}$.}
    \label{fig:Rc_curlB}
\end{figure}

At smaller radii, where $\alpha_b$ increases towards smaller radius (the inner part of the ``V''), the collapse speed increases, and the magnetic field can no longer remain straight. The resulting pinching of the field lines strengthens the magnetic tension force, thereby restoring the magnetic contribution to the force balance. To quantitatively illustrate the origin of the ``V'' shape in $\alpha_{b, R}$ due to a shift of magnetic field geometry, we quantify the magnetic field curvature by defining a curvature radius
\begin{equation}
    R_c \equiv \frac{|\mathbfit{B}|}{[\nabla\times\mathbfit{B}]_R},
    \label{equ:Rc}
\end{equation}
where $|\mathbfit{B}|$ is the magnetic-field strength and $[\nabla\times\mathbfit{B}]_R$ is the radial component of the curl. Figure~\ref{fig:Rc_curlB} shows $R_c$ in the M0.0AD2.0 model. At large radii ($R \gtrsim 400$ au), $R_c$ remains $\gtrsim 50$ au, indicating a roughly parallel magnetic field without an increase in pinching (the outer part of the ``V''). Around $R \sim 250$ au, $R_c$ decreases rapidly to $\lesssim 25$ au, marking a sharp increase in field-line curvature (the inner part of the ``V''). This transition occurs at the same location where $\alpha_b$ reaches its minimum. Consequently, the onset of strong pinching naturally produces the characteristic ``V-shaped'' radial profile of the magnetic term.

\begin{figure}
    \centering
    \includegraphics[width=\linewidth]{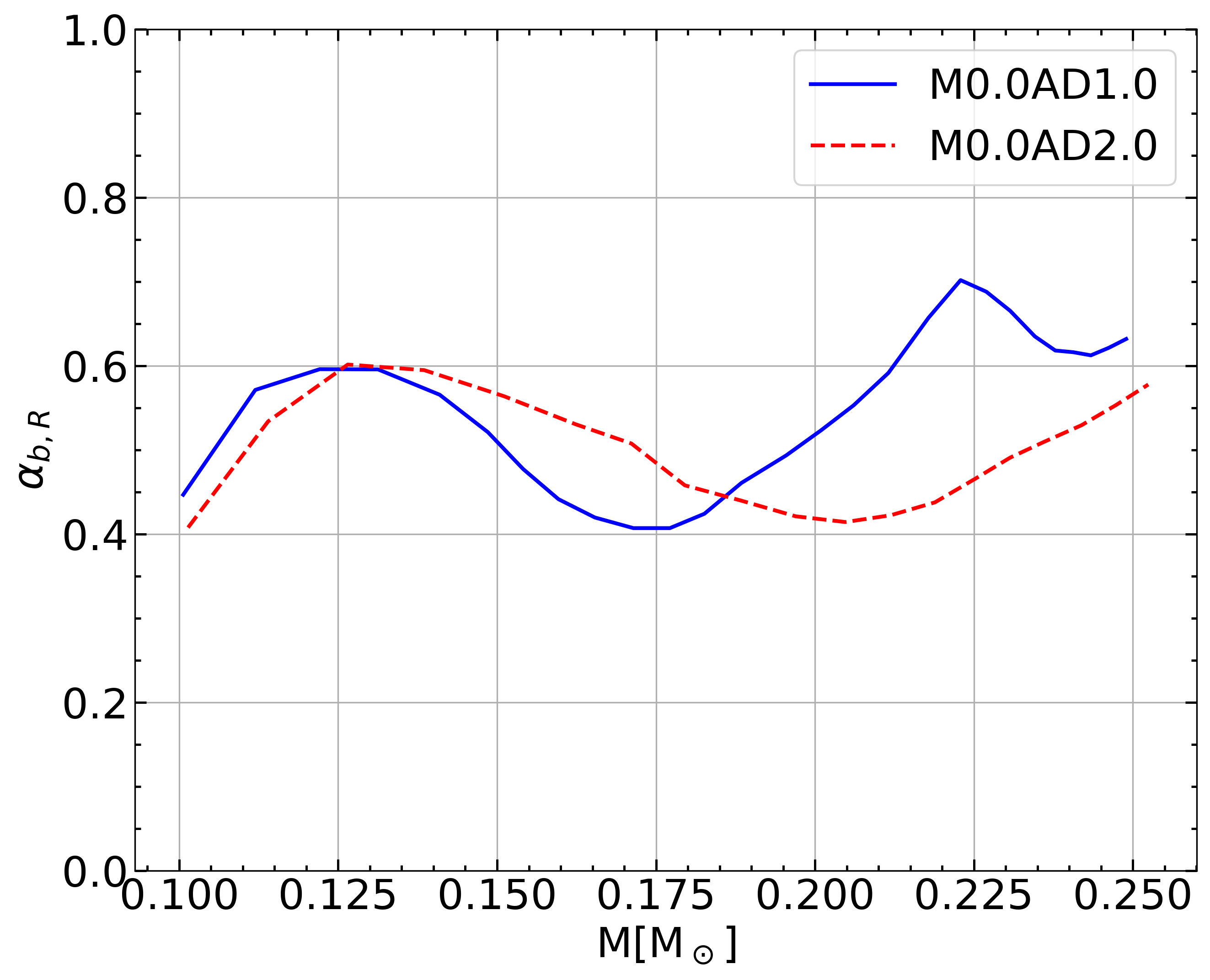}
    \caption{Time evolution of $\alpha_{b,R}$ in the M0.0AD1.0 and M0.0AD2.0 models. At each time, values are averaged over radii between 300 and 600 au. The characteristic ``V'' shape in time arises from the underlying ``V-shaped'' radial profile (see Section~\ref{sec:est_alpha_b_direct}), produced by the effective temporal convolution of this profile as it propagates outward with time.}
    \label{fig:alpha_b_time}
\end{figure}

Acknowledging that there is an expected spatial dependence of the magnetic force contribution, 
we show in Figure~\ref{fig:alpha_b_time} the averaged $\alpha_b$ over the pseudodisk between 300 and 600 au as a function of time. Since this ``V-shaped'' radial profile migrates outward with time, taking a time average at a fixed spatial location is effectively equivalent to convolving this V-shaped structure over time, which preserves the V-shaped form when calculating the time dependence. Nevertheless, the resulting values remain close to $\alpha_b \approx 0.5$ in both models.

\subsection{Estimate $\alpha_{b, R}$ using kinetic information}
\label{sec:est_alpha_b_alpha_v}
A direct measurement of magnetic properties, including magnetic force,  magnetic field strength, and magnetic geometry, is extremely challenging observationally. Kinematic information, available through Doppler shift, is much more easily accessible. In this section, we show that $\alpha_b$ can be estimated using the kinematic information.

With assumptions 1 and 2 quantified in Section~\ref{sec:assumption12}, we can write equation~\ref{equ:mom_equ_simp_01} as
\begin{equation}
    \alpha_{b, R}= 1-\alpha_{t, R}-\alpha_{p, R}-\alpha_{v, R}\approx 0.8-\alpha_{v, R},
\end{equation}
using $\alpha_{t, R}+\alpha_{p, R}\approx0.2$. Since the two ``direct'' ways of measuring $\alpha_{b, R}$ in Section~\ref{sec:est_alpha_b_direct} yield largely the same result (compare $\alpha_{b,1}$ and $\alpha_{b,2}$ in Figure~\ref{fig:alpha_b_measure_direct}), we compare the values measured here using kinematic information with those in method 2 ($\alpha_{b,2}$, i.e., equation~\ref{equ:alpha_b_uplow}), which is what we will use eventually to estimate the magnetic field strength.

\begin{figure*}
    \centering
    \includegraphics[width=\linewidth]{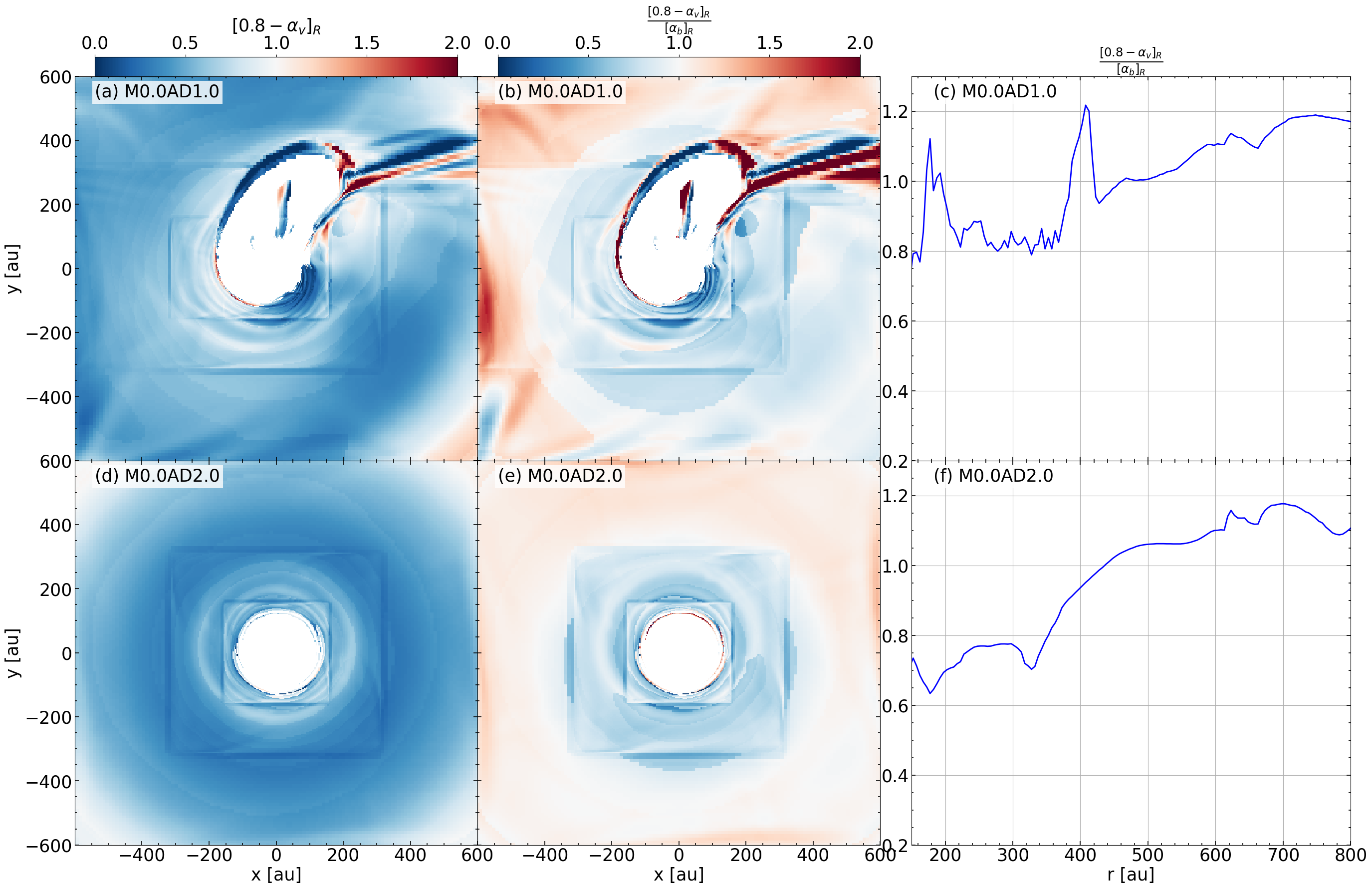}
    \caption{Ratio between the estimated $\alpha_{b,R}$ inferred from gas kinematics ($\alpha_{b,R} = 0.8 - \alpha_{v,R}$) and the actual value measured directly from the simulation using Method~2 (equation~\ref{equ:alpha_b_uplow}; see section~\ref{sec:est_alpha_b_direct}). \textbf{Panel (a)} shows the spatial distribution of $0.8-\alpha_{v,R}$ on the pseudodisk. \textbf{Panel (b)} presents the ratio between the kinematically inferred value and $\alpha_b$ measured from equation~\ref{equ:alpha_b_uplow}. \textbf{Panel (c)} shows the azimuthally averaged ratio from panel (b), demonstrating overall good agreement between the kinematic estimate and the true value. \textbf{Panels (d–f)} show the same quantities as panels (a–c), but for the M0.0AD2.0 model.}
    \label{fig:alpha_v_AD1}
\end{figure*}

Figure~\ref{fig:alpha_v_AD1} shows $0.8 - \alpha_{v, R}$ in panel (a). The inferred values are very similar to those measured directly in Figure~\ref{fig:alpha_b_measure_direct}(b), except near the edge of the magnetic bubble, where the gas kinematics undergo an abrupt transition, as expected. Figure~\ref{fig:alpha_v_AD1}(b) shows the ratio between the estimated and measured $\alpha_{b, R}$, which remains close to unity with only modest spatial variations.

To quantify the level of agreement, Figure~\ref{fig:alpha_v_AD1}(c) presents the column-density--weighted, annulus-averaged ratio. Within $\sim 400$~au, the ratio is slightly below unity, primarily due to the kinematic transition at the magnetic bubble boundary. Beyond $\sim 400$~au, the ratio remains close to 1. Overall, the estimation error remains within $\sim 20\%$ of the directly measured value at all radii.

Similar behavior is found in the M0.0AD2.0 model, shown in Figure~\ref{fig:alpha_v_AD1}(d-f). In this case, there is a systematic underestimate at small radii ($\lesssim 350$~au), primarily caused by a slight overestimate of $\alpha_{b,2}$ relative to $\alpha_{b,1}$ (see Figure~\ref{fig:alpha_b_measure_direct}[f]). As a result, equation~\ref{equ:alpha_b_uplow} slightly overestimates $\alpha_{b, R}$. At larger radii, where $\alpha_{b,2} \approx \alpha_{b,1}$, the ratio between $0.8 - \alpha_{v, R}$ and $\alpha_{b, R}$ converges toward unity.

These results demonstrate that, overall, the value of $\alpha_{b, R}$ can be reliably recovered from $\alpha_{v, R}$, which can be measured using gas kinematics observations. In practice, a useful approximation could be to assume $\alpha_{b, R}$ is roughly constant over space, and we discuss this in section~\ref{sec:sub-freefall envelope}. In the following subsection, we connect this inferred magnetic force ratio to the underlying magnetic field strength.

\subsection{Field geometry}
\label{sec:geometry}
Our method so far only estimates the value of the product of $B_z$ and $B_R$, which is the magnetic stress. This product can be used to obtain a lower limit on the total magnetic field strength. From equation~\ref{equ:alpha_b_uplow}, if we assume that the magnetic field strengths above and below the disk surface are comparable, i.e.,
\begin{equation}
    (B_z/B_R)_\mathrm{above} \approx (B_z/ B_R)_\mathrm{below} \equiv (B_z/ B_R)_\mathrm{surf},
\end{equation}
where the subscript ``surf'' denotes quantities evaluated at the pseudodisk surface. Mathematically, equation~\ref{equ:alpha_b_uplow} can be rearranged to give
\begin{equation}
    B_{\mathrm{pol},\,\mathrm{surf}}^2 = (B_z^2 + B_R^2)_\mathrm{surf} \ge 2 (B_z B_R)_\mathrm{surf} \approx \alpha_{b,R}~4\pi g_R \Sigma .
    \label{equ:Bpol_upperlimit}
\end{equation}
Because $B_z$ and $B_R$ are evaluated at the disk surface, the right-hand side provides a lower limit on the total poloidal magnetic field strength at the pseudodisk surface. We have verified that equation~\ref{equ:Bpol_upperlimit} holds in our simulation, and typically underestimates the magnetic field strength by a factor of a few.

In practice, however, a more relevant quantity is the vertical magnetic flux threading the pseudodisk, which ultimately determines how much magnetic flux is transported into the circumstellar disk. This flux is primarily set by the vertical field component $B_z$. The relationship between $B_R$ and $B_z$ at the pseudodisk surface is geometric and depends on the pinching angle of the magnetic field lines (denoted by $\theta_\mathrm{above}$ and $\theta_\mathrm{below}$ in Figure~\ref{fig:sketch}; $|B_z|\approx |B_R|\tan\theta$; and $\gamma_{zR}\equiv\tan(\theta)$). In principle, this geometry could be constrained observationally through high-resolution dust polarization measurements that resolve the pseudodisk surface \citep[as done in, e.g.,][but on a larger scale]{Choi2025}. However, such observations are challenging and costly.

Alternatively, we can provide an approximate estimate of this geometric parameter using our numerical simulations. We therefore parameterize the field geometry by defining the geometric parameter as 
\begin{equation}
    \gamma_{zR}\equiv\tan\theta\approx(|B_z/B_R|)_\mathrm{surf}
    \label{equ:gamma_zR}
\end{equation}
which allows us to estimate the vertical field strength with equation~\ref{equ:estBz_M0.0}.
Since $B_z$ is approximately vertically uniform within the pseudodisk, this estimate applies throughout the pseudodisk and is not limited to the surface.

Figure~\ref{fig:gamma_plot}(a) shows the geometry parameter $\gamma_{zR}$ in the M0.0AD1.0 model, computing the averaged $B_z/B_R$ at the upper and lower pseudodisk surface. The value stays around $\approx 0.4$ at all locations (Figure~\ref{fig:gamma_plot}[c]). A similar value is obtained in the M0.0AD2.0 model, as shown in Figure~\ref{fig:gamma_plot}(b) and (c). This shows that in the majority of the pseudodisk, albeit there are some radial dependencies, $\gamma_{zR}\approx0.4$, corresponding to a pinching angle $\theta\approx20^\circ$.

\subsection{Estimate field strength}
\label{sec:est_field_strength}
With the geometry factor specified, we are now ready to test equation~\ref{equ:estBz_M0.0}. We compare the estimated magnetic field strength, computed from the right-hand side of equation~\ref{equ:estBz_M0.0}, with the actual value measured in the simulation. The latter is defined as the volume-weighted average of $B_z^2$ within the pseudodisk, denoted $\bar{B}_z^2$. We quantify the agreement using the ratio
\begin{equation}
    \mathcal{R}_\mathrm{est}^2\equiv\frac{\gamma_{zR} \alpha_{b, R}2\pi g_R \Sigma}{\bar{B}_z^2},
    \label{equ:Bz_est_ratio}
\end{equation}
so that the quantity $\mathcal{R}_\mathrm{est}$ reflects the accuracy of magnetic field strength estimation directly. 


Although the most accurate estimation of magnetic field strength ($\mathcal{R}_\mathrm{est}\approx 1$) would be obtained if 2D distributions of $B_z$, $\alpha_{b, R}$, and $\gamma_{zR}$ on the pseudodisk were used, such information is not realistically obtainable observationally. A more meaningful test, therefore, employs simplified, approximate quantities and evaluates the resulting accuracy.

\begin{figure*}
    \centering
    \includegraphics[width=\linewidth]{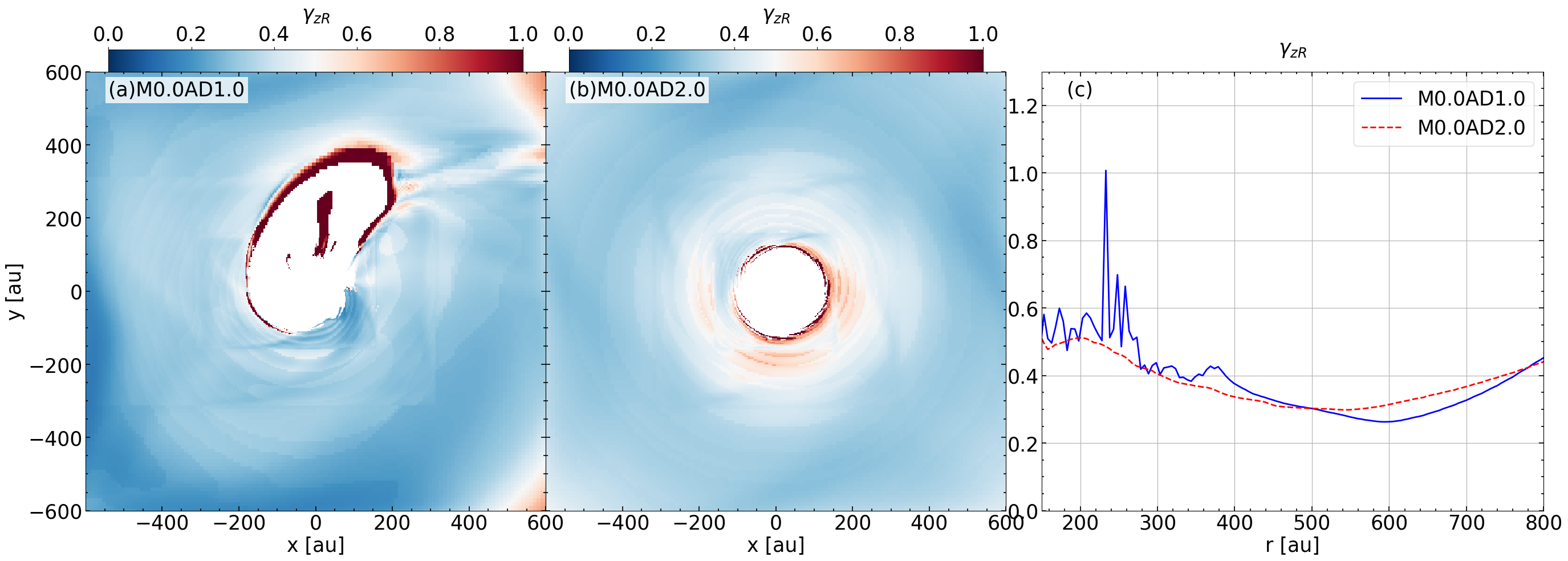}
    \caption{Measurement of $\gamma_{zR}$ in our models. The value shown is the average taken between the upper and lower surfaces of the pseudodisk. \textbf{Panel (a)} shows the M0.0AD1.0 model; \textbf{panel (b)} shows the M0.0AD2.0 model, and \textbf{Panel (c)} shows the azimuthally averaged value to further quantify $\gamma_{zR}$ in both models.}
    \label{fig:gamma_plot}
\end{figure*}

\begin{figure*}
    \centering
    \includegraphics[width=\linewidth]{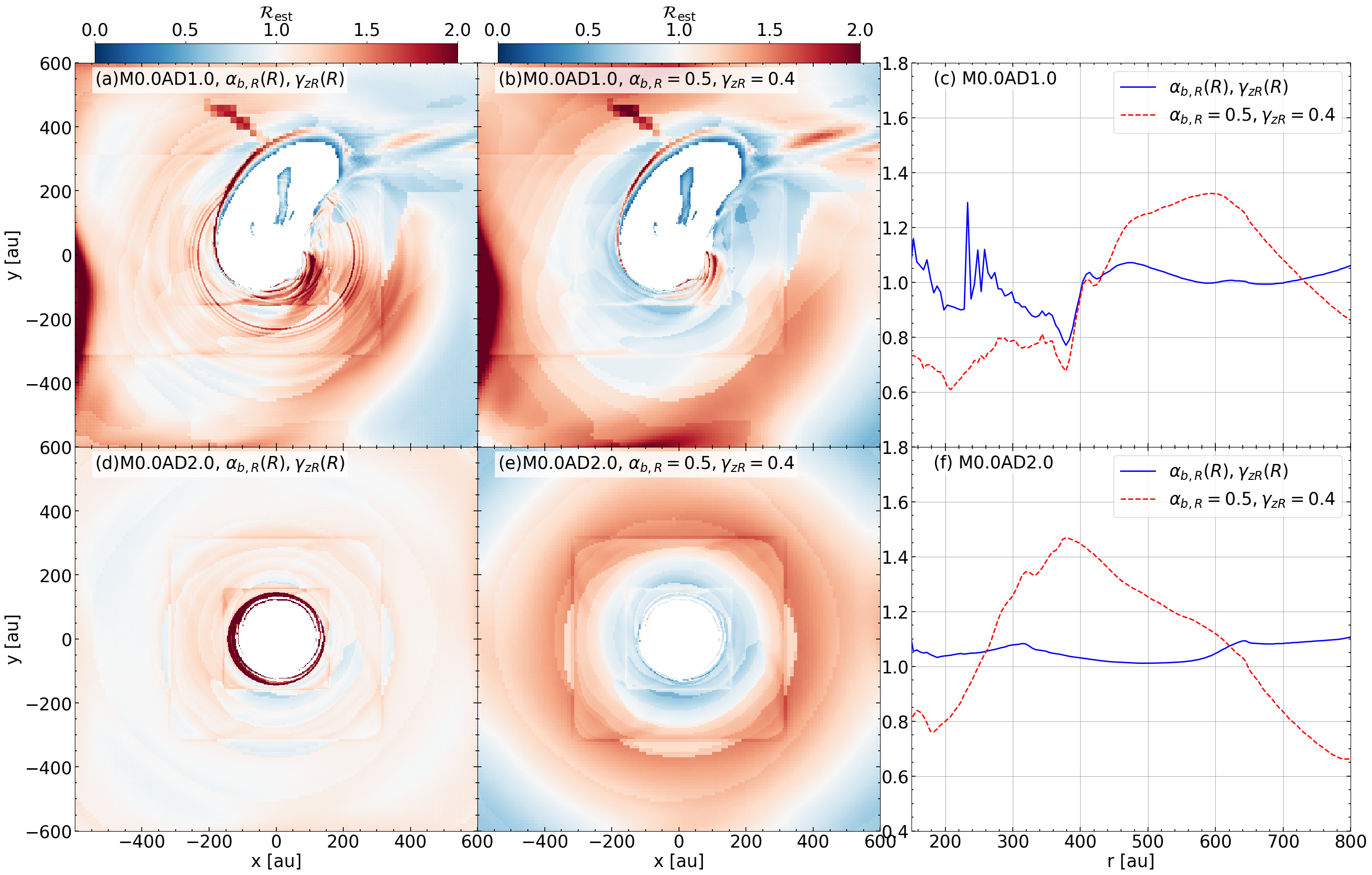}
    \caption{Comparison between the actual vertical magnetic field strength in the simulation, defined as the vertically-averaged $B_z^2$, with the estimation using equation~\ref{equ:estBz_M0.0}. The ratio is defined as $\mathcal{R}$ in equation~\ref{equ:Bz_est_ratio}. The upper panels show the M0.0AD1.0 model and the lower panels show the M0.0AD2.0 model. \textbf{Panels a and d} show the ratio using radially-dependent $\alpha_{b,R}(R)$ and $\gamma_{zR}(R)$ (i.e., taken from the rightmost panels of figure~\ref{fig:alpha_b_measure_direct} and \ref{fig:gamma_plot}). \textbf{Panels b and e} further relax the radial dependence, taking $\alpha_{b,R}=0.5$ and $\gamma_{zR}=0.4$ as constants throughout the pseudodisk. \textbf{Panels c and f} show the azimuthal average of the ratio obtained by either method, quantifying the good agreement between our estimation and the actual value.}
    \label{fig:final_ratio_plot}
\end{figure*}

We first retain only the radial dependence and neglect the $\phi$ dependence, using the 1D profiles $\alpha_{b, R}(R)$ and $\gamma_{zR}(R)$ shown in Figures~\ref{fig:alpha_b_measure_direct}(c), (f) and \ref{fig:gamma_plot}(c), respectively. The resulting $\mathcal{R}_\mathrm{est}$ is shown in Figure~\ref{fig:final_ratio_plot}(a) and (d) for the M0.0AD1.0 and M0.0AD2.0 models, respectively, with their annuli average in panels (c) and (f). The agreement is excellent: the estimated field strength remains within $\sim20\%$ of the true value over most of the pseudodisk.

Motivated by this result, we further simplify the estimation by assuming spatially constant values for both $\alpha_{b, R}$ and $\gamma_{zR}$. Some degradation in accuracy is expected, as both quantities exhibit some level of radial variation (Figures~\ref{fig:alpha_b_measure_direct} and \ref{fig:gamma_plot}). The resulting ratio is shown in Figure~\ref{fig:final_ratio_plot}(b) and (e) for the M0.0AD1.0 and M0.0AD2.0, respectively, with the annuli averaged profiles again shown in panels (c) and (f). As anticipated, the radial ``V-shaped'' feature in $\alpha_{b, R}$ is reflected in $\mathcal{R}_\mathrm{est}$. Nevertheless, the estimated values remain within a factor of $\lesssim40\%$ over most radii, particularly in regions further away from the magnetic bubble. This demonstrates that even highly simplified estimates can recover the magnetic field strength with reasonably good accuracy. 

Similar results are obtained in other snapshots of both simulations, demonstrating that our findings are robust over the majority of the early stages of star formation. The consistency of these results across two models with different ambipolar diffusivities further indicates that our method is applicable to a broad range of astrophysical systems, even when their ionization states are uncertain, as the inferred quantities show only weak sensitivity to the level of ambipolar diffusion in our models.

The estimation of $B_z$ is most directly relevant in star-forming clouds, as it controls the amount of magnetic flux transported into the disk and thus ultimately sets the level of disk magnetization. The total magnetic field strength, $|B|=\sqrt{B^2}$, however, is often of greater interest in an observational context, especially when the pseudodisk is not fully resolved, so we also provide an estimate of the total field strength here.

Unlike $B_z$, which is approximately vertically uniform, the total magnetic field strength is not uniform across the pseudodisk and its surrounding envelope because of magnetic field pinching. As illustrated in Figure~\ref{fig:sketch}, the magnetic field magnitude reaches a local minimum on the pseudodisk, where both the radial and azimuthal components ($B_R$ and $B_\phi$) vanish, leaving only the vertical component $B_z$. Above and below the pseudodisk, the magnetic field strength is enhanced relative to the pseudodisk midplane. We thus estimate the magnetic field strength there.

To estimate the total magnetic field strength, we need to estimate both $B_R$ and $B_\phi$. The former, $B_R$, is directly available through the definition of $\gamma_{zR}$. To estimate the latter, $B_\phi$, we can define
\begin{equation}
    \gamma_{z\phi}\equiv |B_z/B_\phi|_\mathrm{surf}
    \label{equ:gamma_zphi}
\end{equation}
at the pseudodisk surface, analogous to the definition of $\gamma_{zR}$, and estimate its value in appendix~\ref{app:gamma_zphi}. The total magnetic field strength is thus given by
\begin{equation}
    |B|_\mathrm{surface}^2=\Big(1+\frac{1}{\gamma_{zR}^2}+\frac{1}{\gamma_{z\phi}^2}\Big)|B_z|^2\approx\Big(1+\frac{1}{\gamma_{zR}^2}\Big)|B_z|^2,
    \label{equ:est_Bsq}
\end{equation}
where the subscript ``surface'' highlights that this magnetic field strength is estimated at the surface of the pseudodisk. The last approximation in equation~\ref{equ:est_Bsq} comes from $\gamma_{z\phi}>\gamma_{zR}$, as $B_\phi$ has a relatively small magnitude compared with $B_z$ and $B_R$. This is expected as the protostellar envelope is dominated by radial infall motion instead of rotation. We have verified that using equation~\ref{equ:est_Bsq}, the estimation of the total magnetic field strength in both M0.0AD1.0 and M0.0AD2.0 models yields reasonably good results with similar accuracy as estimating $B_z$ alone. This shows that our method can be extended to estimate total magnetic field strength requiring minimal additional information.

We briefly summarize our method for estimating the magnetic field strength. Using equation~\ref{equ:estBz_M0.0}, measurements of the gravitational acceleration ($g_R$) and the local column density of the pseudodisk ($\Sigma$), together with canonical values of $\alpha_{b, R} = 0.5$ and $\gamma_{zR} = 0.4$, allow a rough estimate of the pseudodisk magnetic field strength with an accuracy of $\sim 40\%$. This accuracy can be further improved if $\alpha_{b, R}$ is constrained observationally through gas kinematics (Section~\ref{sec:est_alpha_b_alpha_v}), or if $\gamma_{zR}$ is inferred from polarization observations that probe the field geometry (Section~\ref{sec:geometry}). The total magnetic field strength at the pseudodisk surface can be roughly recovered by using the definition of $\gamma_{zR}$ with equation~\ref{equ:est_Bsq}.

\section{Extend to Models with Initial Turbulence}
\label{sec:validation_turb}
We have verified our method and estimated the $\alpha_{b, R}$ and $\gamma_{zR}$ parameters in the models with an initial laminar protostellar core. In reality, most observed protostellar cores exhibit some level of turbulence, making it important to determine how well our method applies in a more dynamical system.

\citet[][]{Tu2024a} have shown that the complexity of the structure of a protostellar envelope increases with the initial turbulence strength. Specifically, turbulence distorts the pseudodisk, which lies predominantly on the equatorial plane, into fully 3D, sheetlet-like structures that do not follow any specific axis or orientation. Consequently, measuring the magnetic field ``above'' and ``below'' each sheetlet, which involves identifying the ``perpendicular'' direction of the sheetlet at every location, becomes a challenging and impractical task. We thus take a more empirical approach in testing our method in the turbulent models: as we have justified and tested our method in the laminar models extensively, and that the physics of the sheetlets are analogous to those of the pseudodisk \citep[see][]{Tu2024a}, it is reasonable to presume that the method should work reasonably well in the turbulent model with a similar set of parameters. Thus, we directly test this statement and verify the parameter choice.

We test equation~\ref{equ:est_Bperp} by comparing the estimated magnetic field strength with the magnetic field strength measured directly in the simulation, defined in several different ways. We choose the $\hat{z}$ direction as the integration direction for the column density ($\Sigma_z$), such that the verification is carried out in the $\hat{R}$–$\hat{\phi}$ plane of a cylindrical coordinate. The gravitational acceleration $g_s$ is taken to be the cylindrical-$\hat{R}$ component of the gravitational force evaluated at the midplane. This choice is also motivated by observational considerations: while the spherical-radial direction is difficult to infer observationally, the cylindrical-radial direction is much more readily accessible. The force-balance parameter is estimated as $\alpha_b\approx0.3$ using the free-fall speed estimation \citep[see figure 5d in][and section~\ref{sec:sub-freefall envelope}]{Tu2024a}; since the system is much more dynamic in the turbulent model, the exact magnetic field geometry relation between $B_z$ and $B_R$ in this case would be different from those in the non-turbulent model. Therefore, instead of interpreting $\gamma_{zR}$ as the field geometry parameter, it should be interpreted as an empirical correction factor. For simplicity, we take it as our canonical value $\gamma_{zR}=0.4$. 

We compare our estimate with the magnetic field strength defined in three different ways, corresponding to three distinct physical interpretations. The first definition uses the column-weighted vector-averaged $B_z$, accounting for the possibility that oppositely directed magnetic fields along the line of sight may partially cancel. The corresponding ratio, defined similarly as equation~\ref{equ:Bz_est_ratio}, is
\begin{equation}
    \mathcal{R}^2_\mathrm{vec} \equiv \frac{\gamma_{zR}\alpha_{b,R} 2\pi g_s\Sigma_z}{\langle B_z\rangle^2}.
    \label{equ:turb_est_Rvec}
\end{equation}

The second definition uses the column-weighted magnetic field strength in the line-of-sight direction, i.e., the column-weighted $B_z^2$,
\begin{equation}
    \mathcal{R}^2_\mathrm{str} \equiv \frac{\gamma_{zR}\alpha_{b,R} 2\pi g_s\Sigma_z}{\langle B_z^2\rangle}.
    \label{equ:turb_est_Rstr}
\end{equation}

Finally, we compare our estimate with the total magnetic field strength by considering the magnitude of the full magnetic field vector. In the turbulent model, $\gamma_{zR}$ is no longer interpreted strictly as the geometric ratio between $B_z$ and $B_R$, but rather as an effective empirical correction factor that encapsulates the complex three-dimensional field structure. Nevertheless, we apply equation~\ref{equ:est_Bsq} directly to assess how accurately our method recovers the overall magnetic field strength when used without modification. This comparison provides a stringent test of the robustness and practical applicability of our prescription in the limit of a fully dynamical, turbulent environment. The estimator is defined as
\begin{equation}
    \mathcal{R}^2_\mathrm{tot} \equiv \frac{\big(1+\frac{1}{\gamma_{zR}^2}\big)\gamma_{zR}\alpha_{b,R} 2\pi g_s\Sigma_z}{\langle \boldsymbol{B}^2\rangle}.
    \label{equ:turb_est_Rtot}
\end{equation}

\begin{figure*}
    \centering
    \includegraphics[width=\linewidth]{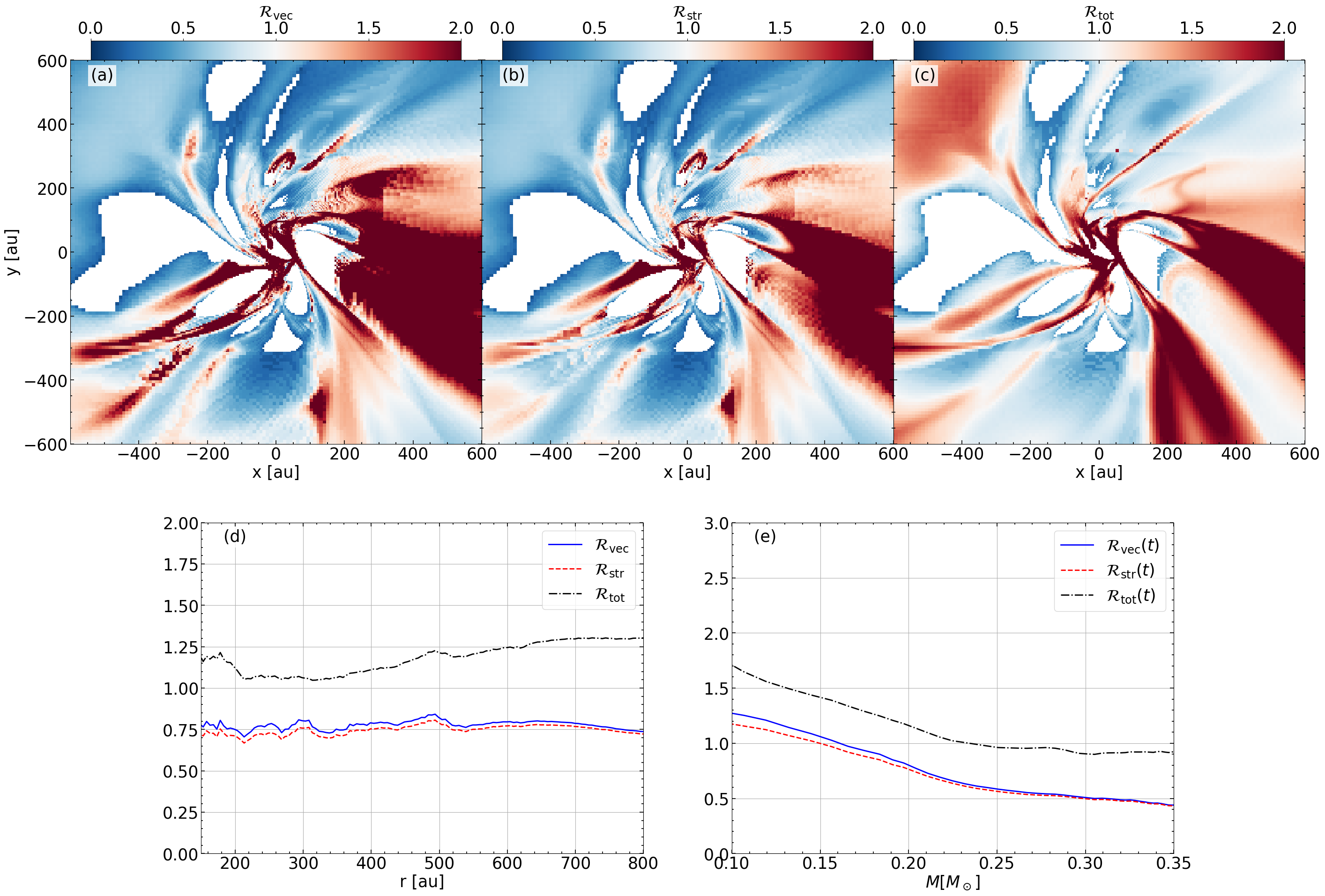}
    \caption{Comparison between the values in the turbulent model \citep[M1.0AD1.0 in][]{Tu2024a} and magnetic field strength estimation through direct application of our method. \textbf{The upper panels (panels a, b, and c)} show 2D agreement using the definitions in equation~\ref{equ:turb_est_Rvec}, \ref{equ:turb_est_Rstr}, and \ref{equ:turb_est_Rtot}, respectively. \textbf{Panel d} shows the azimuthal-average of the agreement parameter, and \textbf{panel e} shows the radial and azimuthal averaged agreement parameters over time, illustrating that our method can be applied well in turbulent models.}
    \label{fig:turb_est_B}
\end{figure*}

Figures~\ref{fig:turb_est_B}[a–c] show the two-dimensional distributions of $\mathcal{R}_\mathrm{vec}$, $\mathcal{R}_\mathrm{str}$, and $\mathcal{R}_\mathrm{tot}$ at $M_\star = 0.2~M_\odot$, respectively, for Model M1.0AD1.0 with an initial turbulent Mach number of one. Their azimuthally averaged profiles are shown in Figure~\ref{fig:turb_est_B}[d]. The first estimator and second estimator, quantified by $\mathcal{R}_\mathrm{vec}$ and $\mathcal{R}_\mathrm{str}$ respectively, show similar and good agreement, remaining within a factor of two almost everywhere in the two-dimensional maps. Their azimuthally averaged profiles show a modest systematic offset, typically underestimating the field strength by a factor of about $25\%$, but without strong radial dependence. Owing to this weak spatial trend, we further examine the time evolution of the domain-averaged ratios in Figure~\ref{fig:turb_est_B}(e). While some time dependence is evident--most notably a gradual decrease of the ratio values as the stellar mass grows--the deviation remains within a factor of two until $M_\star \approx 0.3~M_\odot$ (about 60\% of the core mass).

The third estimator, $\mathcal{R}_\mathrm{tot}$, which targets the total magnetic field strength, shows a similar level of deviations. In contrast to the first two estimators, it generally overestimates the field strength by a factor of $\lesssim 1.5$. This level of agreement persists throughout the evolution, remaining within a factor of two at all times.

The time-dependence in both the $B_z$ estimates and the total field strength estimates shows a similar trend: at earlier times, the method tends to yield a larger estimation of the magnetic field strength, whereas at later times it tends to yield a lower value. This behavior likely reflects uncertainties associated with estimating the column density via straight line-of-sight integration, as well as possible time dependence in the parameters $\alpha_{b,R}$ and $\gamma_{zR}$. Nevertheless, throughout the protostellar collapse phase, the estimated magnetic field generally remains within a factor of $\sim 2$ of the actual magnetic field strength in each method. Overall, our method therefore provides a robust estimate of the magnetic field strength over a wide range of evolutionary stages.


\section{Discussion}
\label{sec:discussion}
\subsection{Sub-free-fall envelope: a shortcut to estimate $\alpha_{b,R}$}
\label{sec:sub-freefall envelope}
Using our numerical simulations, we find that although some radial variation is present, the parameter $\alpha_{b,R}$ (Equation~\ref{equ:alpha_b}) remains close to $\alpha_{b,R} \simeq 0.5$ in both the M0.0AD1.0 and M0.0AD2.0 models. Since $\alpha_b$ represents the ratio of magnetic force to gravitational force, this implies that roughly half of the gravitational force is counterbalanced by magnetic stresses. This highlights the crucial role of magnetic fields in significantly retarding the infall of the protostellar envelope.

This result allows a further physical interpretation. If the magnetic contribution to force balance is approximately similar at different radii, then $\alpha_{b,R}$ can also be interpreted as the fractional reduction of the infall speed relative to free fall. Similar interpretation can be applied to $\alpha_t$ and $\alpha_p$ (equation~\ref{equ:mhd alpha equ}). The radial equation of motion can be written as
\begin{equation}
    \frac{d^2 r}{dt^2} = (1-\alpha_t-\alpha_p-\alpha_b) \frac{GM}{r^2}\approx(0.8-\alpha_b)\frac{GM}{r^2},
\end{equation}
where we used $\alpha_t+\alpha_p\approx0.2$ (section~\ref{sec:assumption12}).
This expression is formally identical to the classical free-fall equation, except for the factor $(0.8-\alpha_b)$. Physically, this factor acts as an effective softening of the stellar gravity, analogous to a reduced stellar mass. Because the functional form of the equation of motion is unchanged, the usual relationship between free-fall velocity and stellar mass still applies. Using the fact that spherical-$\hat{r}$ and cylindrical-$\hat{R}$ coincide on the pseudodisk midplane, we can approximate
\begin{equation}
    v_r\approx v_R = (0.8-\alpha_{b, R})^{1/2} v_{\mathrm{ff}} .
\end{equation}
Thus, if the local free-fall speed can be estimated (e.g., from the stellar mass) and the radial velocity is measured observationally (e.g., via Doppler shifts), $\alpha_{b, R}$ can be inferred directly from kinematic information. We have verified that this approximation recovers $\alpha_{b, R}$ reasonably well in our numerical simulation.

The fact that the magnetic force contributes significantly in our model is consistent with other theoretical modeling \citep[e.g.,][]{Tomida2015, Lam2019, Hirano2025, Wang2025}, who also identify sub-free velocity in (Radiative) MHD simulations. Additionally, our finding that magnetic forces substantially reduce infall speeds is consistent with observational evidence for sub-free-fall collapse (see, for example, L1157 below). \citet[][]{Mottram2013} find that slower-than-free-fall motions are required to explain observations of IRS4A (see also \citealt{Su2019}), potentially indicating magnetic retardation of infall in that system. 

\subsection{Connection to the DCF method and the KTH method}
\label{sec:dcf_koch}
One widely used technique for estimating magnetic field strength is the Davis–Chandrasekhar–Fermi (DCF) method, originally proposed by \citet{Davis1951} and \citet{Chandrasekhar1953}. The method relies on two key assumptions. First, the magnetic field is relatively ordered and uniform, allowing it to be decomposed into a large-scale component and a turbulent component, with the latter being a fraction $\sigma$ of the total field ($\delta B = \sigma B$). Second, it assumes approximate energy equipartition between turbulent kinetic and turbulent magnetic energies, $\delta E_K \approx \delta E_B$. Under these assumptions, the turbulent kinetic energy can be inferred from spectral line broadening, while the turbulent magnetic fraction $\sigma$ can be estimated from dust polarization measurements, together enabling an estimate of the total magnetic field strength.

The DCF method is generally not well-suited for estimating magnetic field strength in an actively collapsing protostellar envelope. In such environments, both underlying assumptions of the method may break down. The gas and magnetic field possess preferred directions of motion as they are drawn inward by gravity, leading to a highly non-uniform field configuration that tends to concentrate toward the collapse center rather than remaining ordered with small perturbations. Moreover, the strong, systematic infall motions imply that turbulent kinetic and magnetic energies are unlikely to be in equipartition, further undermining the validity of the DCF assumptions and making magnetic field estimates from this method highly uncertain in collapsing regions.

Another method that has been proposed to estimate magnetic field strength is the ``KTH method'' \citep{Koch2012}, which shares a similar physical foundation with our approach. Both methods are derived from the MHD momentum equation (equation~\ref{equ:mhd momentum equation}), but differ in their assumptions and simplifications. The KTH method projects the magnetic tension force and the ``gas'' force, defined as the combined gravitational and pressure-gradient forces, onto the same plane-of-sky direction and assumes local force balance. The gas force is inferred from the intensity gradient of the emission, and, by estimating a curvature radius of the magnetic field (analogous to our equation~\ref{equ:Rc}), the magnetic field strength can then be obtained from the magnetic tension term.

The primary difference between our method and the KTH method lies in the treatment of the gas advection term and the time-dependence term in the momentum equation. In the KTH method, both terms are necessarily neglected, and only magnetic, gravitational, and pressure forces are considered. However, as shown in sections~\ref{sec:assumption12} and \ref{sec:est_alpha_b_alpha_v}, the time-dependence term contributes approximately $10\%$ of the total force balance ($\alpha_t\approx0.1$), while the advection term ($\alpha_v$) can contribute up to $\sim40\%$. Neglecting these terms may therefore lead the KTH method to overestimate the magnetic force contribution and, consequently, the inferred magnetic field strength.

In addition, although both methods share a similar physical interpretation of polarization observations and their implications for magnetic forces, the manner in which these forces are applied differs. In the KTH method, magnetic tension is treated as a large-scale force, allowing the curvature inferred from large-scale dust polarization observations to be directly associated with the magnetic field curvature. In contrast, in our method, the magnetic tension force is more localized, acting primarily within the thin pseudodisk. As a result, the relevant magnetic field curvature (estimated in equation~\ref{equ:Rc} and Figure~\ref{fig:Rc_curlB}) is also localized to the pseudodisk. Consequently, curvature estimates derived from polarization observations that do not resolve the pseudodisk may underestimate the true magnetic field curvature and should be regarded as lower limits.

Finally, our method and the KTH method differ in how forces are projected. The KTH method projects forces onto the plane of the sky from the observer's perspective, whereas we project forces along the spherical-radial direction measured from the system center. In a collapsing protostellar system, where gravity dominates and acts primarily in the radial direction, the projected gas-motion direction (inferred from intensity gradients in the KTH method) and the magnetic curvature direction (from polarization observations) are expected to align approximately with the gravitational force. Under these circumstances, the different projection schemes may be approximately equivalent, suggesting that the two approaches can be used in a complementary manner.

\subsection{Application to L1157}
\label{sec:est_L1157}
In this section, we present a proof-of-concept application and estimate the magnetic field strength in L1157 using available literature data. We emphasize that this application is intended as an illustrative demonstration and a conceptual test of the method rather than a precise measurement.

The L1157 is a very young protostellar system that hosts a relatively low-mass star \citep[][]{Looney2007} and a relatively massive and flattened envelope \citep[$1.1M_\odot$, ][]{Beltran2004}. The magnetic field strength has been estimated in \citet[][]{Stephens2013} using both the DCF and KTH method, yielding a field strength of $1\sim3~\mathrm{mG}$. These estimates allow us to apply our method and compare it with both previous methods using available observational constraints.

To apply equation~\ref{equ:estBz_M0.0} (the low-turbulent version of equation~\ref{equ:est_Bperp} and \ref{equ:est_Bperp_nosq}), we need an estimate of the central stellar mass (to estimate $g_R$); an estimate of the column density of the pseudodisk $\Sigma$, and some estimates of $\alpha_{b,R}$ and $\gamma_{zR}$. We take the central stellar mass as $M_\star\approx0.2M_\odot$, following \citet[][]{Looney2007} and \citet[][]{Sharma2020} with some uncertainties. The density of the pseudodisk $\Sigma$ is not directly measured in the literature, so some estimates are needed. Using the best fit range in \citet[][]{Looney2007}, the mass density is $\rho\sim\mathrm{a\ few}\times 10^{-18}~\mathrm{g~cm^{-3}}$, and given the envelope is observed to be somewhat large in thickness, we take 
\begin{equation}
    \Sigma_\mathrm{L1157}\sim0.5\rho r\sim1.35\times10^{-2}~\mathrm{g~cm^{-2}},
\end{equation}
where $r=300~\mathrm{au}$ is derived from the resolution limit and the reference distance used in \citet{Looney2007}, but accounting for the uncertainty in distance measurement \citep[][]{Sharma2020}.

Substantial sub-free fall velocity is observed in L1157 \citep[][]{Kristensen2012}, motivating us to take a somewhat large estimate of $\alpha_{b,R}\sim0.7$ for our calculations. There is no observational constraint on the geometric factor $\gamma_{zR}$, so we take our canonical value of $\gamma_{zR}=0.4$. Using these values, we estimate
\begin{equation}
    \begin{split}
        |B^2_{z, \mathrm{L1157}}|=\big(1.77\times10^{-4}~\mathrm{G}\big)^2&\Big(\frac{\alpha_{b,R}}{0.7}\Big)\Big(\frac{\gamma_{zR}}{0.4}\Big) \\
        &\Big(\frac{\Sigma}{\Sigma_\mathrm{L1157}}\Big)\Big(\frac{GM}{G~0.2M_\odot}\Big)\Big(\frac{r}{3\times10^2~\mathrm{au}}\Big)^{-2}.
    \end{split}
\end{equation}
Note that this estimated magnetic field strength is only the vertical component, and it can help to yield an estimation of the magnetic flux into the disk. To estimate the total magnetic field strength, we need to estimate $B_R$ and $B_\phi$, both of which are readily available through $\gamma_{zR}$ (section~\ref{sec:geometry}). We can estimate the total magnetic field strength at $300~\mathrm{au}$ as
\begin{equation}
    |B|_\mathrm{\mathrm{300~au}}=\Big(1+\frac{1}{\gamma_{zR}^2}\Big)^{1/2}|B_z|_\mathrm{\mathrm{300~au}}\sim 5\times10^{-4}~\mathrm{G}. 
\end{equation}
Note that this estimation has a spatial dependence of the magnetic field strength: assuming the column-density stays roughly constant as a function of radius, at $r=100~\mathrm{au}$, the magnetic field strength would be about $1.5\times10^{-3}~\mathrm{G}$. Acknowledging some uncertainties due to this spatial dependence, this estimation is somewhat comparable to the estimated value by the DCF method in \citet{Stephens2013}, and is smaller than the values estimated through the KTH method, which could be an overestimate (see section~\ref{sec:dcf_koch}). Follow-up high-resolution Zeeman observation of L1157 can be helpful to test our method in practice.

\section{Conclusion and Summary}
\label{sec:conclusion}
We present a new method for estimating both the vertical magnetic field strength and the total magnetic field strength in pseudodisk- or sheetlet-dominated collapsing protostellar envelopes. Our method is derived from the MHD momentum equation, and its underlying assumptions, approximations, and key parameters are verified and calibrated using non-ideal MHD numerical simulations. The main results can be summarized as follows:

\begin{enumerate}
    \item The application of our method of estimating the vertical magnetic field strength ($|B_\mathrm{\perp}|=(\gamma_{\perp\parallel}~\alpha_{b, s}~2\pi g_s\Sigma)^{1/2}$, equation~\ref{equ:est_Bperp}; and equation~\ref{equ:estBz_M0.0} in the case of a equatorial pseudodisk) requires only two observationally accessible quantities: the projected gravitational acceleration toward the center of collapse, $g_s$, and the face-on column density of the pseudodisk or sheetlets, $\Sigma$.
    \item The two dimensionless parameters appearing in equation~\ref{equ:est_Bperp}--the magnetic fraction of the total force, $\alpha_b \approx 0.5$, and the geometric field parameter, $\gamma_{\perp\parallel} \approx 0.4$--remain close to these canonical values with little variation in space or time in our simulations. This relative constancy allows magnetic field strength estimates in the absence of detailed gas kinematics or high-resolution polarization data.
    \item Using information available through observations, the parameter $\alpha_b$ can be approximated using gas kinematics; if the magnetic contribution to retard infall stays roughly constant spatially and temporally, the ratio between the observed gas velocity and the inferred free-fall velocity can be used as an approximation for estimating $\alpha_b$. $\gamma_{\perp\parallel}$ can be constrained through high-resolution polarization observations that probe the magnetic field geometry near the pseudodisk surface.
    \item Our method is applicable to both turbulent and non-turbulent protostellar envelopes and is insensitive to the ambipolar diffusion coefficient, which depends on the highly uncertain local ionization rate. This makes the method robust across a wide range of physical conditions.
    \item While our approach shares a similar physical foundation with the KTH method \citep[][]{Koch2012}, we show using numerical simulations that the gas advection term plays a significant role in the force balance. Neglecting this term can lead to an overestimation of the magnetic field strength.
    \item We apply our method to L1157 using gravitational acceleration and column density estimates from the literature. Our inferred magnetic field strength is broadly consistent with previous studies.
\end{enumerate}

\section*{Acknowledgements}

YT is supported by NASA 80NSSC24K1285, and acknowledges support from an interdisciplinary fellowship from the University of Virginia. ZYL is supported in part by NASA 80NSSC20K0533, NSF AST-2307199, JWST-GO-02104.002-A, JWST-GO-08872.003-A, and the Virginia Institute of Theoretical Astronomy (VITA). Computing resources were provided by the NASA High-End Computing (HEC) Program through the NASA Advanced Supercomputing (NAS) Division at Ames Research Center, NSF computing allocation PHY-250048, and the Research Computing (RIVANNA and AFTON supercomputers) at the University of Virginia.

\section*{Data Availability}

The data underlying this article will be shared on reasonable request to the corresponding author.



\bibliographystyle{mnras}
\bibliography{example} 




\appendix
\section{Overview of M0.0AD2.0 model}
\label{app:M0.0AD2.0}
In this appendix, we provide an overview of the M0.0AD2.0 model, which was not presented in \citet{Tu2024a}.

\begin{figure*}
    \centering
    \includegraphics[width=\linewidth]{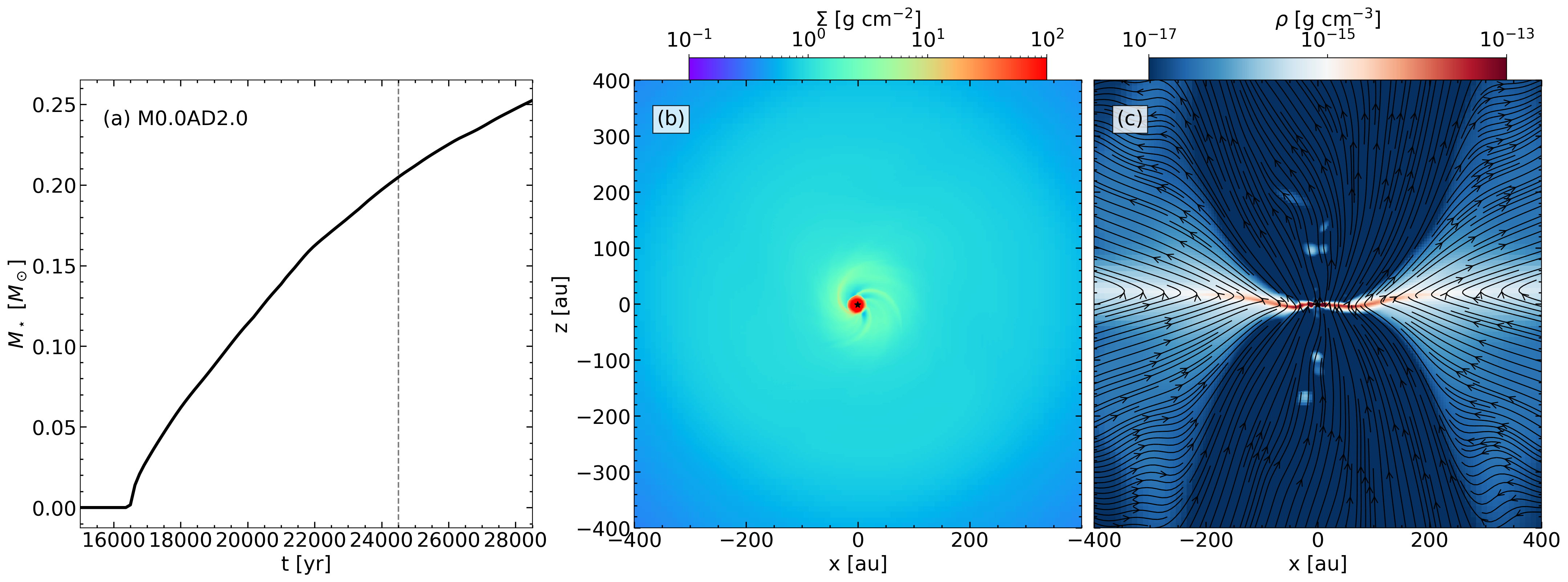}
    \caption{Overview of the M0.0AD2.0 model. \textbf{Panel (a)} shows the stellar (sink particle) mass as a function of time; \textbf{panel (b)} shows the face-on column density at the representative time when $M_\star=0.2 ~M_\odot$, and \textbf{panel (c)} shows a meridian slice of density, overplotted by magnetic field lines.}
    \label{fig:AD2.0_overview}
\end{figure*}

The initial conditions of the M0.0AD2.0 model are identical to those of the M0.0AD1.0 model, except that the ambipolar diffusivity coefficient is increased by a factor of two, corresponding to an assumed cosmic-ray ionization rate lower by a factor of four. Figure~\ref{fig:AD2.0_overview}(a) shows the evolution of the stellar mass; the vertical dashed line marks the representative epoch used throughout this paper, at which $M_\star\approx0.2~M_\odot$. Figure~\ref{fig:AD2.0_overview}(b) shows the face-on column density at this time. A relatively small disk, with a radius of $\sim20~\mathrm{au}$, is surrounded by the pseudodisk, which occupies the remainder of the plotted region. In contrast to the M0.0AD1.0 model—where a prominent magnetic bubble is present at $M_\star\approx0.2~M_\odot$—the pseudodisk in the M0.0AD2.0 model remains largely axisymmetric. This difference is likely due to the higher magnetic diffusivity, which reduces magnetic flux accumulation near the central region and delays the formation of a magnetic bubble.

To illustrate the pseudodisk geometry, Figure~\ref{fig:AD2.0_overview}(c) shows a density slice through the meridional plane, overplotted with magnetic field lines. The pseudodisk exhibits a slight wobble about the midplane, caused by weak turbulence driven by the early outflow at the time of stellar formation. Nevertheless, it remains well confined to the midplane throughout the simulation. The magnetic field lines display a characteristic pinched morphology similar to that seen in the M0.0AD1.0 and M1.0AD1.0 models presented in \citet{Tu2024a}.

\section{The azimuthal direction}
\label{app:gamma_zphi}
The validation and estimation presented in section~\ref{sec:validation_noturb} focus on the magnetic field components in the $\hat{z}$ and cylindrical-$\hat{R}$ directions. In this appendix, we demonstrate that the azimuthal magnetic field component, $B_\phi$, is small compared to both $B_z$ and $B_R$, and therefore does not significantly affect our total field-strength estimates.

\begin{figure*}
    \centering
    \includegraphics[width=\linewidth]{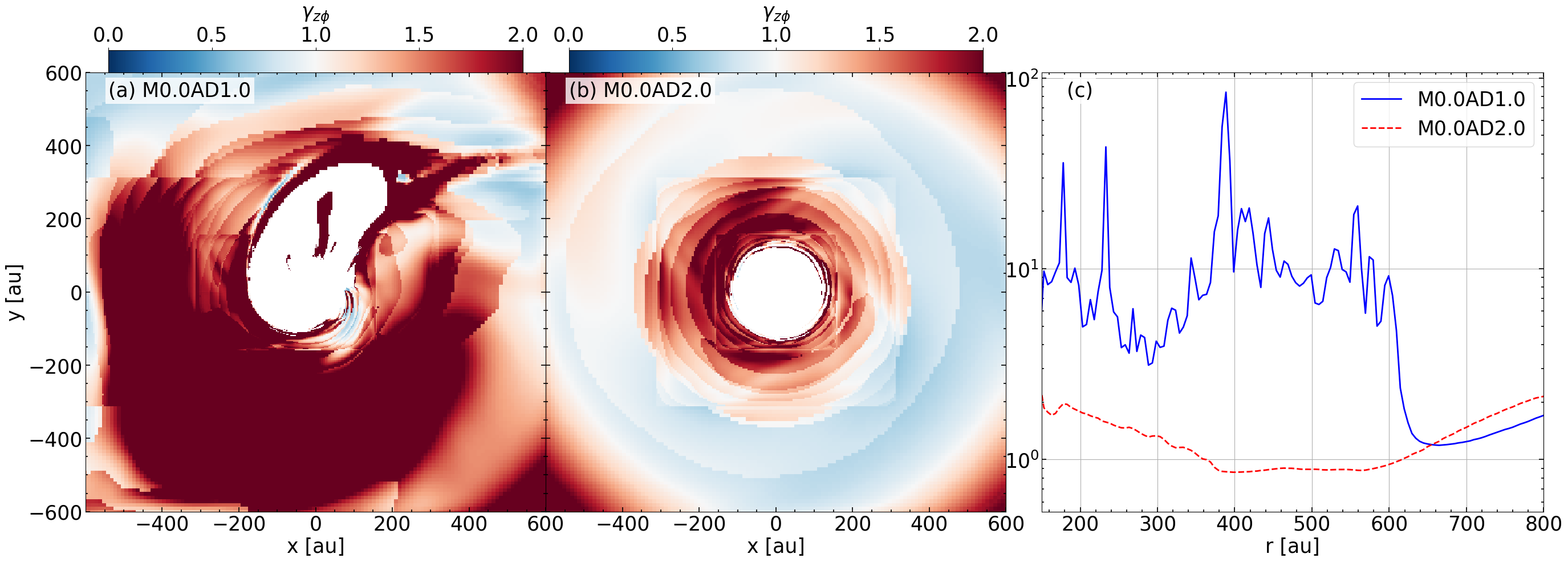}
    \caption{$\gamma_{z\phi}$ (equation~\ref{equ:gamma_zphi}) in the M0.0AD1.0 model (panel a) and M0.0AD2.0 model (panel b). Panel c shows their azimuthal average, indicating $\gamma_{z\phi}\gtrsim1$.}
    \label{fig:gamma_zphi}
\end{figure*}

We focus on $B_\phi$ evaluated at the surface of the pseudodisk, as discussed in section~\ref{sec:est_field_strength}. Figure~\ref{fig:gamma_zphi} shows the ratio $\gamma_{z\phi}$ (defined in equation~\ref{equ:gamma_zphi}) for the M0.0AD1.0 and M0.0AD2.0 models at the representative epoch when $M_\star = 0.2,M_\odot$. Despite some spatial variation, $\gamma_{z\phi}$ remains generally greater than unity throughout the pseudodisk. In particular, over the bulk of the pseudodisk ($\lesssim 600~\mathrm{au}$), the ratio is typically of order a few to $\sim100$, indicating that $B_\phi$ is subdominant relative to $B_z$.

We have verified that $\gamma_{z\phi}\gtrsim 1$ throughout most of the simulation, especially at early times. At later stages, as vertical magnetic flux accumulates toward the center, this ratio increases but remains modest, reaching values of order unity or a few. Similar behavior is observed in the M0.0AD2.0 model, although $\gamma_{z\phi}$ can be slightly smaller at the outer pseudodisk (still $\gtrsim 0.8$), before increasing at later times as the system evolves.

Given that the radial magnetic field component is almost always stronger than the vertical component in our simulations ($\gamma_{zR} \approx 0.4$), the contribution of $B_\phi$ to the total magnetic field strength is subdominant. As a result, variations in $\gamma_{z\phi}$ do not significantly affect our estimate of the total magnetic field strength (equation~\ref{equ:est_Bsq}), and the azimuthal component can be safely neglected or treated as a minor correction in our analysis.
\bsp	
\label{lastpage}
\end{document}